   \documentstyle[psfig]{l-aa}
\def \etal     {et al.}
\def \ie       {i.\,e.}
\def \eg       {e.\,g.}
\def \vLSR     {\hbox{${v_{\rm LSR}}$}}
\def \delv     {\hbox{$\Delta v_{1/2}$}}
\def \TMB      {\hbox{$T_{\rm MB}$}}
\def \Tkin     {\hbox{$T_{\rm kin}$}}
\def \Tsys     {\hbox{$T_{\rm sys}$}}
\def \Mgas     {\hbox{$M_{\rm gas}$}}                     
\def \Msol     {\hbox{M$_{\odot}$}}                       
\def \Mvir     {\hbox{$M_{\rm vir}$}}                     
\def \Egrav    {\hbox{$E_{\rm grav}$}}                    
\def \Eturb    {\hbox{$E_{\rm turb}$}}                    
\def \Etherm   {\hbox{$E_{\rm therm}$}}                   
\def \HII      {H\,{\sc ii}}
\def \Halpha   {H$\alpha$}
\def \kms      {\hbox{${\rm km\,s}^{-1}$}}                
\def \Kkms     {\hbox{${\rm K\,km\,s}^{-1}$}}             
\def \percc    {\hbox{${\rm cm}^{-3}$}}                   
\def \arcdeg   {\hbox{$^{\circ}$}}                        
\def \RA#1     {\hbox{$\alpha_{#1}$}}                     
\def \Dec#1    {\hbox{$\delta_{#1}$}}                     
\def \MOLH     {\hbox{H$_2$}}                             
\def \HCC      {\hbox{HC$_2$}}                            
\def \twCO     {\hbox{$^{12}$CO}}                         
\def \thCO     {\hbox{$^{13}$CO}}                         
\def \CseO     {\hbox{C$^{17}$O}}                         
\def \CeiO     {\hbox{C$^{18}$O}}                         
\def \WATER    {\hbox{H$_2$O}}                            
\def \METH     {\hbox{CH$_3$OH}}                          
\def \CYCP     {\hbox{C$_3$H$_2$}}                        
\def \HCOp     {\hbox{HCO$^+$}}                           
\def \Cp       {\hbox{C$^+$}}                             
\def \OO       {\hbox{O$_2$}}                             
\def \C#1      {\hbox{$^{#1}$C}}                          
\def \O#1      {\hbox{$^{#1}$O}}                          
\def \ISOC     {\hbox{$^{12}$C$/^{13}$C}}                 
\newcommand{\see}[1]{$^{\rm #1)}$}
\newcommand{\vol}[1]{{\rm #1}}
\def \AaA      {{\rm A\&A}}
\def \AaAS     {{\rm A\&AS}}
\def \AaAR     {{\rm A\&AR}}
\def \ApJ      {{\rm ApJ}}
\def \ApJS     {{\rm ApJS}}
\def \MNRAS    {{\rm MNRAS}}
\def \IAU      {{\rm IAU Symposium}}
\def \PASA     {{\rm Proc. Astron. Soc. Aust.}}
\def \JPCRef   {{\rm J. Phys. Chem. Ref. Data}}
\def \bra#1    {$\left\{\makebox{\rule[-#1ex]{0pt}{#1ex}}\right.$}
\def \ket#1    {$\left.\makebox{\rule[-#1ex]{0pt}{#1ex}}\right\}$}
\def \uspace#1 {\makebox{\rule[#1ex]{0pt}{2ex}}}
\def \dspace   {\makebox{\rule[-2ex]{0pt}{2ex}}}

\begin{document}

\thesaurus{11(09.01.1; 09.13.2; 11.01.1; 11.09.4; 11.13.1; 13.19.3)}

\title{Molecular abundances in the Magellanic Clouds}

\subtitle{I. A multiline study of five cloud cores {
       \thanks{Based on observations with the Swedish-ESO Submillimeter
               Telescope (SEST) at the European Southern Observatory (ESO),
               La Silla, Chile} } }

\author{
   Y.-N.~Chin\inst{1}
   \thanks{{\it Present address:\/} Institute of Astronomy and Astrophysics,
                                    Academica Sinica, P.O.Box 1-87, Nankang,
                                    Taipei, Taiwan},
   C.~Henkel\inst{2}, J.B.~Whiteoak\inst{3}, T.J.~Millar\inst{4},
   M.R.~Hunt\inst{5}, \and C.~Lemme\inst{2}
   \thanks{{\it Present address:\/} Institut f\"ur Planetenerkundung,
                                    DLR, Rudower Chaussee 5,
                                    D-12489 Berlin, Germany}
}

\offprints{Y.-N.~Chin}

\institute{
   Radioastronomisches Institut der Universit\"at Bonn,
   Auf dem H\"ugel 71, D-53121 Bonn, Germany
\and
   Max-Planck-Institut f\"ur Radioastronomie,
   Auf dem H\"ugel 69, D-53121 Bonn, Germany
\and
   Paul Wild Observatory, Australia Telescope National Facility, CSIRO,
   Locked Bag 194, Narrabri NSW 2390, Australia
\and
   Department of Physics, UMIST, P O Box 88, Manchester M60 1QD, United Kingdom
\and
   University of Western Sydney Nepean, P.O. Box 10, Kingswood, NSW 2747,
   Australia
}

\date{Received date ; accepted date}

\maketitle

\begin{abstract}

   Nine \HII\ regions of the LMC were mapped in \thCO(1--0) and
   three in \twCO(1--0) to study the physical properties of the
   interstellar medium in the Magellanic Clouds.
   For N113 the molecular core is found to have a peak position which differs
   from that of the associated \HII\ region by 20\arcsec.
   Toward this molecular core the \twCO\ and \thCO\ peak
   \TMB\ line temperatures of 7.3\,K and 1.2\,K are the highest
   so far found in the Magellanic Clouds.
   The molecular concentrations associated with N113, N44BC, N159HW, and N214DE
   in the LMC and LIRS\,36 in the SMC were investigated
   in a variety of molecular species to study the chemical properties
   of the interstellar medium.
   $I$(\HCOp)/$I$(HCN) and $I$(HCN)/$I$(HNC) intensity ratios
   as well as lower limits to the $I$(\thCO)/$I$(\CeiO) ratio
   were derived for the rotational 1--0 transitions.
   Generally, \HCOp\ is stronger than HCN, and HCN is stronger than HNC.
   The high relative \HCOp\ intensities are consistent with a
   high ionization flux from supernovae remnants and young stars,
   possibly coupled with a large extent of the \HCOp\ emission region.
   The bulk of the HCN arises from relatively compact dense cloud cores.
   Warm or shocked gas enhances HCN relative to HNC.
   From chemical model calculations it is predicted that $I$(HCN)/$I$(HNC)
   close to one should be obtained with higher angular resolution
   ($\la$ 30\arcsec) toward the cloud cores.
   Comparing virial masses with those obtained from the integrated CO intensity
   provides an \MOLH\ mass-to-CO luminosity conversion factor of
   $1.8 \times 10^{20}$\,mol\,cm$^{-2}$\,(\Kkms)$^{-1}$ for N113 and
   $2.4 \times 10^{20}$\,mol\,cm$^{-2}$\,(\Kkms)$^{-1}$ for N44BC.
   This is consistent with values derived for the Galactic disk.


\end{abstract}

\markboth{\small Y.-N.~Chin, C.~Henkel, J.B.~Whiteoak, et al.:
                 Molecular abundances in the Magellanic Clouds}
         {\small Y.-N.~Chin, C.~Henkel, J.B.~Whiteoak, et al.:
                 Molecular abundances in the Magellanic Clouds}

\section{Introduction}
\label{sec:MC1-Introduction}

   The Large Magellanic Cloud (LMC) and the Small Magellanic Cloud (SMC),
   the two galaxies closest to the Milky Way, provide unique opportunities to
   study astrophysical processes (\eg\ Westerlund 1991).
   In regard to the interstellar medium, there are three outstanding properties
   which motivate detailed investigations:
   (1) the Magellanic Clouds consist of material characterized by
       a smaller metallicity than the Milky Way;
   (2) their UV radiation fields are stronger than in the solar neighborhood;
   (3) they provide a large number of targets at a well determined distance.
   The first two properties have far reaching consequences for
   the astrophysical conditions of the interstellar medium:
   an absence of dust and extinction in the molecular cloud envelopes results
   in reduced shielding against the intense UV radiation and decreases
   the sizes of molecular clouds relative to atomic gas components
   (\eg\ Lequeux \etal\ 1994).
   The low metallicities are also expected to have
   consequences on molecular abundances.

   The ESO-SEST Key Programme (\eg\ Israel \etal\ 1993) included
   observations of the \twCO(1--0) and \thCO(1--0) spectral lines
   toward IRAS sources and \HII\ regions in the LMC and the SMC.
   With the Australia Telescope Compact Array in Narrabri,
   Hunt \& Whiteoak (1994) discovered a second 4.8\,GHz compact
   radio component in the LMC \HII\ region N159 (hereafter N159HW),
   where no \Halpha\ emission had been found.
   This detection motivated our search for \thCO(1--0) cores offset from
   the centers of some of the most prominent \HII\ regions.
   In order to determine the position of brightest emission in molecular clouds,
   we thus mapped nine \HII\ regions in the LMC,
   which have particularly strong \twCO(1--0) and \thCO(1--0) line intensities
   according to the ESO-SEST Key Programme.
   Toward peaks of three of these sources and N159HW 
   we also made 3\,mm multiline studies to elucidate the chemical and
   physical conditions of the cloud cores.

   The SMC is even more metal poor than the Large Magellanic Cloud.
   Hence a comparison of molecular data from a variety of environments like
   the inner and outer Galactic disk and the LMC also has to include the SMC.
   This widens the range of covered metallicities by a factor of 3.
   In previous studies, the molecular line observations toward the
   SMC were confined to \twCO\ and \thCO\ transitions (Rubio \etal\ 1993a,b).
   We therefore observed LIRS\,36, which shows the brightest \twCO(1--0)
   and \thCO(1--0) emission peaks observed with the ESO-SEST
   Key Programme, in a variety of molecular transitions.

\begin{table}
   \caption[]
           {Positions and velocities for observed objects}
 \label{tbl:MC1-source}
   \begin{flushleft}
   \begin{tabular}{l l r r c}
   \hline
   \uspace1
   Galaxy  & Object   & \multicolumn{1}{c}{\RA1950 }
           &   \multicolumn{1}{c}{\Dec1950 }                   &  \vLSR   \\
   \dspace &          & \multicolumn{1}{c}{[$^{\rm h\ m\ s}$]}
           &   \multicolumn{1}{c}{[\arcdeg\ \arcmin\ \arcsec]} &  [\kms]  \\
   \hline
   \uspace1
   SMC     & LIRS\,36 &   0~44~50.5  & $-$73~22~33  &  126.1   \\
   \uspace1
   LMC     & N79      &   4~52~09.5  & $-$69~28~21  &  233.5   \\
           & N83A     &   4~54~17.0  & $-$69~16~23  &  245.0   \\
           & N11      &   4~55~35.5  & $-$66~38~48  &  279~~   \\
           & N105A    &   5~10~05.6  & $-$68~57~00  &  241.6   \\
           & N113     &   5~13~40.2  & $-$69~25~37  &  234.8   \\
           & N44BC    &   5~22~10.6  & $-$68~00~32  &  283.0   \\
           & N55A     &   5~32~30.0  & $-$66~29~21  &  290~~   \\
           & N159HW   &   5~40~04.4  & $-$69~46~54  &  238.3   \\
           & N160     &   5~40~09.0  & $-$69~40~13  &  238.9   \\
   \dspace & N214DE   &   5~40~36.3  & $-$71~11~30  &  228.9   \\
   \hline
   \end{tabular}
   \end{flushleft}
\ \ \\
\end{table}

\begin{table}
   \caption[]
           {Summary of observed transitions in $\lambda$ $\approx$ 3\,mm range.}
 \label{tbl:MC1-freq}
   \begin{flushleft}
   \begin{tabular}{l @{~} l @{} l @{} r@{.}l @{} c}
   \hline
   \multicolumn{2}{l}{Molecule}      & \uspace1
   & \multicolumn{2}{c}{\hspace{-3.2mm} Frequency}   & Beamwidth \\
   \multicolumn{2}{l}{\& Transition} & \dspace
   & \multicolumn{2}{c}{\hspace{-3.2mm} [GHz]}       & [\arcsec] \\
   \hline
   \uspace1
   \CYCP          & $J$=2$_{1,2}-$1$_{0,1}$  &  &  85&338890 & 61 \\
   \uspace1.5
   \HCC           & \begin{tabular}{l} $N$=1--0, \\
                                       ~~~ $J$=3/2--1/2 \end{tabular}
                  & \bra1 \begin{tabular}{l} $F$=2--1     \\ $F$=1--0
                                             \end{tabular}
     & \multicolumn{2}{c}{\hspace{-3.2mm}
                          \begin{tabular}{r@{.}l}  87&316925 \\
                                   \hspace{0.6mm}  87&328624 \end{tabular}}
                                                             & 60 \\
   \uspace3.5
   HCN            & $J$=1--0
                  & \bra2 \begin{tabular}{l} $F$=1--1     \\ $F$=2--1  \\
                                             $F$=0--1     \end{tabular}
     & \multicolumn{2}{c}{\hspace{-3.2mm}
                          \begin{tabular}{r@{.}l}  88&630416 \\
                                   \hspace{0.6mm}  88&631847 \\
                                                   88&633936 \end{tabular}}
                                                             & 59 \\
   \uspace0.5
   \HCOp          & $J$=1--0                 &  &  89&188518 & 58 \\
   \uspace0.5
   HNC            & $J$=1--0                 &  &  90&663543 & 57 \\
   \uspace0.5
   \METH          & $J$=2--1                 &  &  96&741420 & 54 \\
   \uspace0.5
   CS             & $J$=2--1                 &  &  97&980968 & 53 \\
   \uspace0.5
   \CeiO          & $J$=1--0                 &  & 109&782160 & 47 \\
   \uspace0.5
   \thCO          & $J$=1--0                 &  & 110&201353 & 47 \\
   \uspace0.5
   \CseO          & $J$=1--0                 &  & 112&358780 & 46 \\
   \uspace3.5
   CN             & \begin{tabular}{l} $N$=1--0, \\
                                       ~~~ $J$=3/2--1/2 \end{tabular}
                  & \bra2 \begin{tabular}{l} $F$=3/2--1/2 \\ $F$=5/2--3/2 \\
                                             $F$=1/2--1/2 \end{tabular}
     & \multicolumn{2}{c}{\hspace{-3.2mm}
                          \begin{tabular}{r@{.}l} 113&488140 \\
                                                  113&490982 \\
                                                  113&499639 \end{tabular}}
                                                             & 46 \\
   \uspace0.5 \dspace
   \twCO          & $J$=1--0                 &  & 115&271204 & 45 \\
   \hline
   \end{tabular}
   \end{flushleft}
\end{table}

\begin{figure*}
   \vspace*{-197 mm}
   \hspace*{-3 cm} \psfig{figure=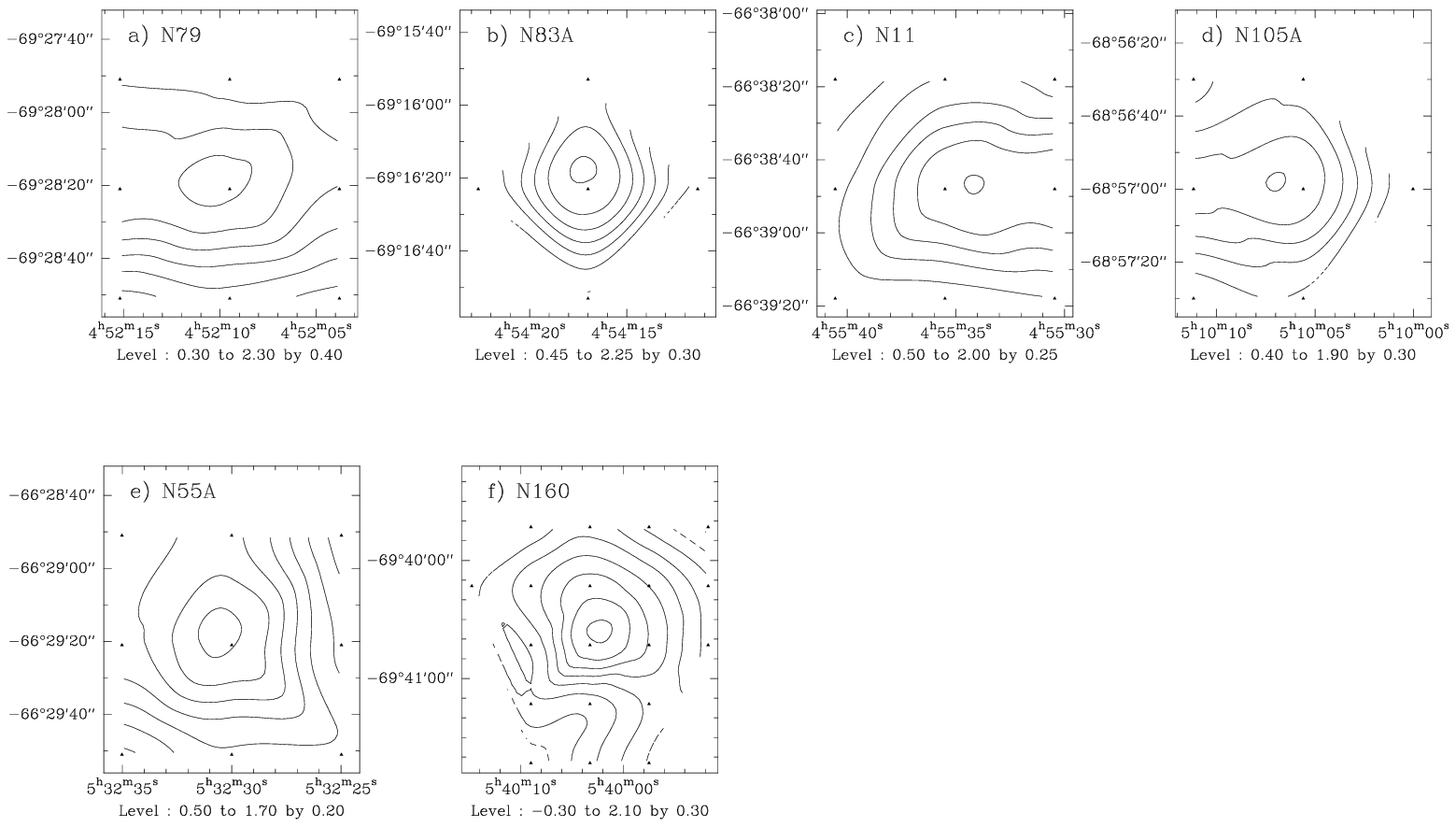,width=23 cm}\\
   \vspace*{-62 mm}
   \begin{flushright} \begin{minipage}{88 mm}
   \caption[]
           {\thCO\ contour maps observed toward
            {\bf a)} N79   (contour levels: 0.30 to 2.30 by 0.40 \Kkms,
                     integration between 239 and 252 \kms\ (\vLSR)),
            {\bf b)} N83A  (contour levels: 0.45 to 2.25 by 0.30 \Kkms,
                     integration between 239 and 252 \kms),
            {\bf c)} N11   (contour levels: 0.50 to 2.00 by 0.25 \Kkms,
                     integration between 274 and 286 \kms),
            {\bf d)} N105A (contour levels: 0.40 to 1.90 by 0.30 \Kkms,
                     integration between 237 and 246 \kms),
            {\bf e)} N55A  (contour levels: 0.50 to 1.70 by 0.20 \Kkms,
                     integration between 282 and 294 \kms),
            {\bf f)} N160  (contour levels: -0.30 to 2.10 by 0.30 \Kkms,
                     integration between 231 and 243 \kms).
            Typical r.m.s.\ values are between 0.1 and 0.2 \Kkms.}
 \label{fig:contour}
   \end{minipage} \end{flushright}
\end{figure*}

\begin{figure}[hbt]
   \vspace*{-24 mm}
   \hspace*{-6 mm} \psfig{figure=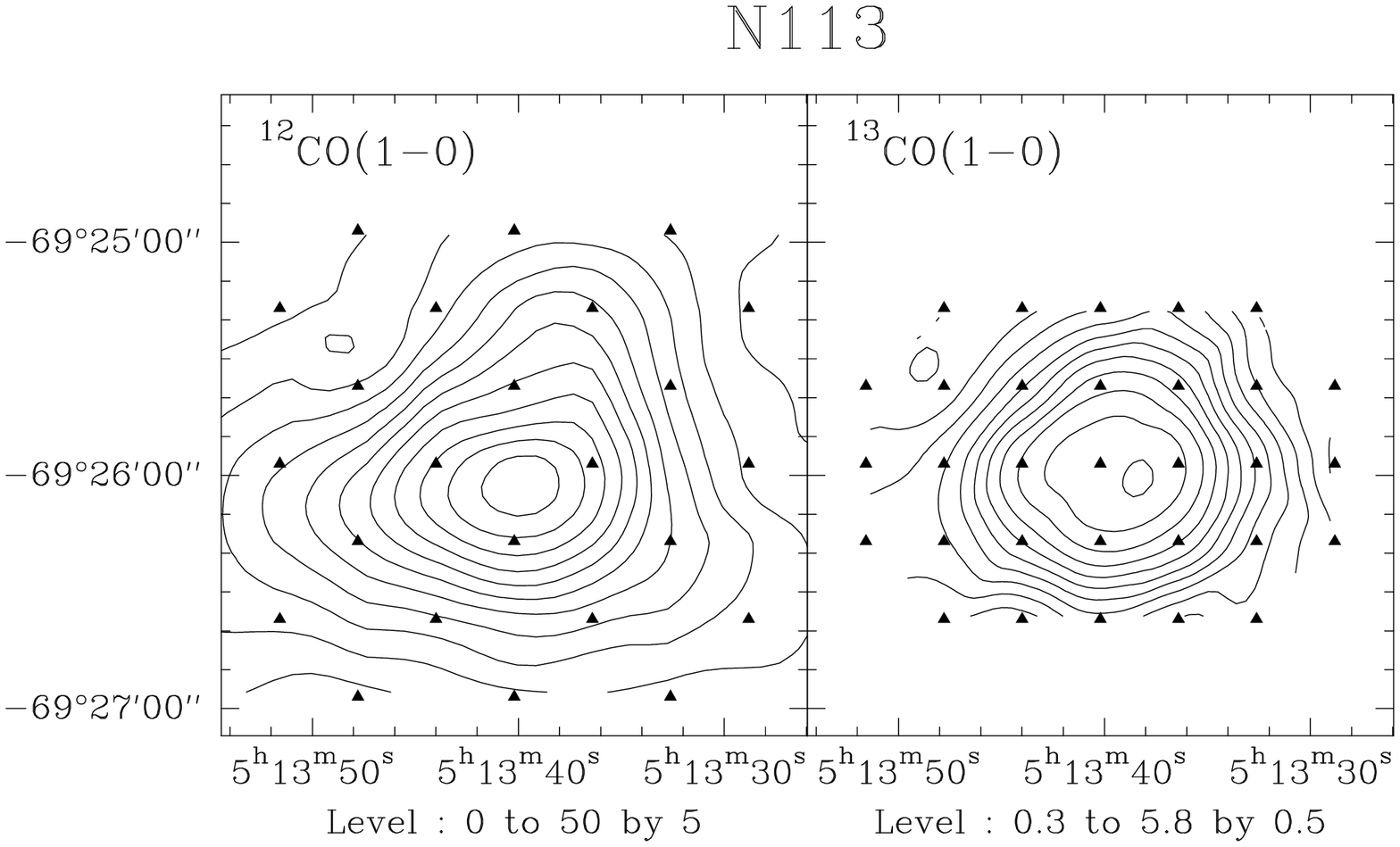,height=7 cm,angle=-90}
   \vspace*{-6 mm}
   \caption[]
           {Contour maps of N113 in
            {\bf a)} \twCO(1--0), where the contour levels are
             0 to 50 by 5 \Kkms\ with the integration between
             230 and 242 \kms\ in \vLSR;
            {\bf b)} \thCO(1--0), where the contour levels are
             0.3 to 5.8 by 0.5 \Kkms\ with the integration between
             \vLSR\ = 231 and 240 \kms.
            Typical r.m.s.\ values are 0.4 and 0.15 \Kkms, respectively.}
 \label{fig:contour-N113}
\end{figure}

\begin{figure}[hbt]
   \vspace*{-24  mm}
   \hspace*{-6 mm} \psfig{figure=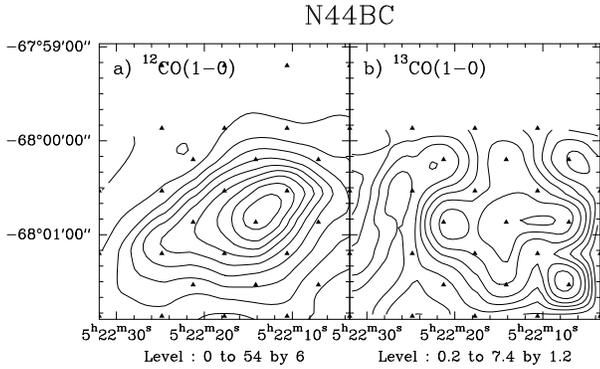,height=7 cm,angle=-90}
   \vspace*{-6 mm}
   \caption[]
           {Contour maps of N44BC in
            {\bf a)} \twCO(1--0), where the contour levels are
             0 to 54 by 6 \Kkms\ with the integration between
             276 and 290 \kms\ in \vLSR;
            {\bf b)} \thCO(1--0), where the contour levels are
             0.4 to 7.6 by 1.2 \Kkms\ with the integration between
             \vLSR\ = 276 and 290 \kms.
            Typical r.m.s.\ values are 0.4 and 0.13 \Kkms, respectively.}
 \label{fig:contour-N44BC}
\end{figure}

\begin{figure}[hbt]
   \vspace*{-10 mm}
   \hspace*{-06 mm} \psfig{figure=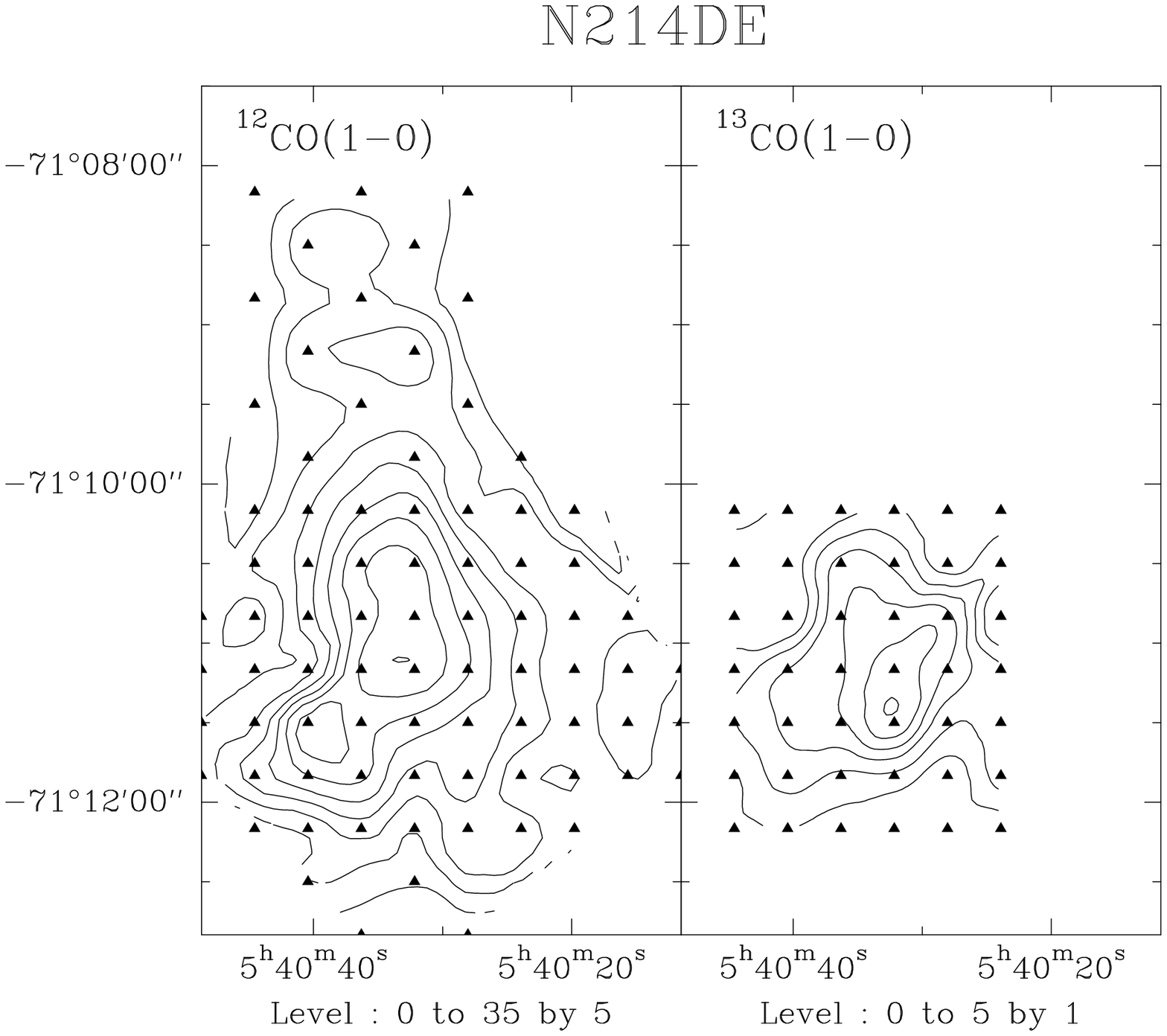,height=7 cm,angle=-90}
   \vspace*{-2 mm}
   \caption[]
           {Contour maps of N214DE in
            {\bf a)} \twCO(1--0), where the contour levels are
             0 to 35 by 5 \Kkms\ with the integration between
             220 and 240 \kms\ in \vLSR;
            {\bf b)} \thCO(1--0), where the contour levels are
             0 to 5 by 1 \Kkms\ with the integration between
             \vLSR\ = 220 and 240 \kms.
            Typical r.m.s.\ values are 0.7 and 0.23 \Kkms, respectively.}
 \label{fig:contour-N214DE}
\end{figure}

\section{Observations}
\label{sec:MC1-Observations}

   Positions and radial velocities of the sources observed are displayed
   in Table\,\ref{tbl:MC1-source}.
   Using the 15-m Swedish-ESO Submillimetre Telescope (SEST) at La Silla, Chile,
   most of the $\lambda$ $\approx$ 3\,mm measurements were carried out in
   September 1993, May 1994 and January 1995.
   Frequencies of observed transitions (taken from Lovas 1992) and
   correspondent antenna beamwidths are summarized in Table\,\ref{tbl:MC1-freq}.
   A Schottky 3-mm receiver yielded overall
   system temperatures (\Tsys), including the sky,
   of order 400\,K on a main-beam brightness temperature (\TMB) scale.
   \Tsys\ was significantly higher only for \twCO(1--0), reaching 800\,K.
   In January 1996 a new SIS receiver was employed with \Tsys\ $\approx$
   500\,K for the $J$=1--0 transitions of \twCO\ and \thCO.
   The backend was an acousto-optical spectrometer with 2000 contiguous
   channels and a channel separation of 43\,kHz
   (0.11 -- 0.15\,\kms\ for frequency range 115 -- 85\,GHz).
   The observations were carried out in a dual beam-switching
   mode (switching frequency 6\,Hz) with a beam throw
   of 11\arcmin 40\arcsec\ in azimuth.
   \twCO(1--0) measurements were also carried out in a frequency-switching mode.
   A comparison of these two sets of data showed consistency in line shapes
   and intensities, confirming that insignificant emission was present
   at the reference positions.
   The on-source integration time of each spectrum varied from
   8 minutes for \twCO\ to 260 minutes for HCN toward LIRS\,36.
   All spectral intensities obtained were transformed to a \TMB\ scale,
   and corrected for a main-beam efficiency 0.74 (Dr.~L.B.G.~Knee priv.\ comm.).
   The pointing accuracy, obtained from measurements of SiO maser sources,
   was better than 10\arcsec.

   The software package CLASS was used for data reduction.
   In most cases a linear baseline correction was sufficient.
   A higher order baseline was occasionally needed for the spectra obtained
   in the frequency-switching mode, but in all cases the spectral lines were
   sufficiently narrow so that baseline removal posed no problems.

\begin{figure*}
   \vspace*{-1 cm}
   \hspace*{0 cm} \psfig{figure=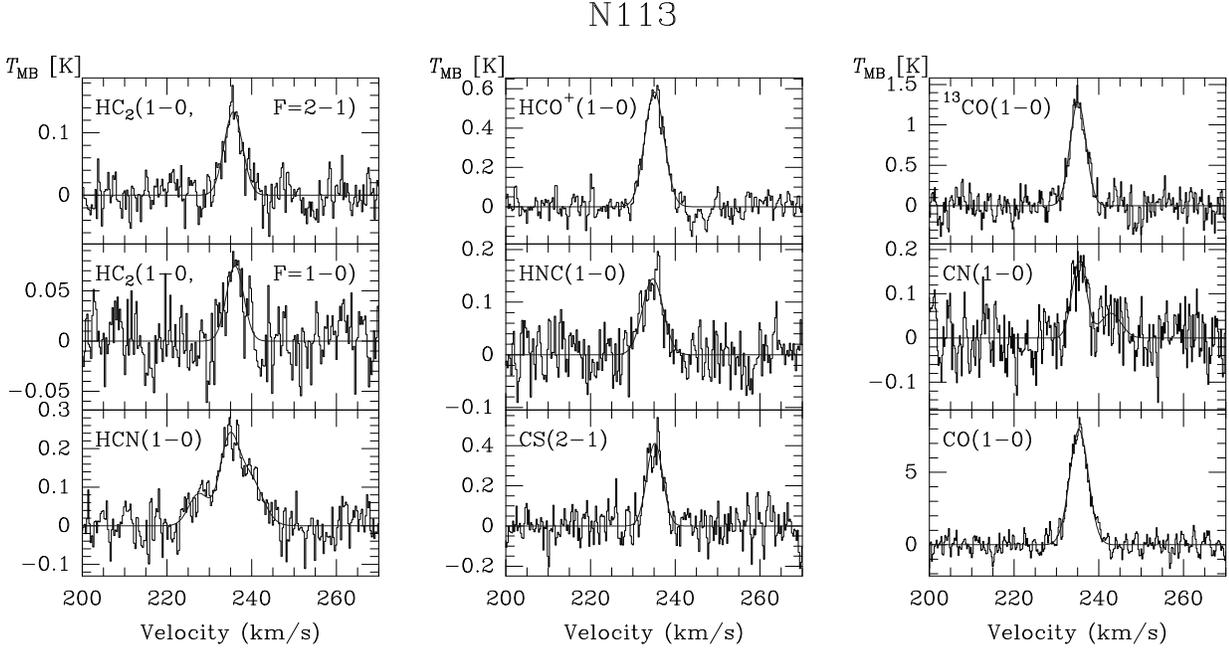,width=18 cm,angle=-90}
   \vspace*{-3 cm}
   \caption[]
           {Molecular spectra toward the core of N113}
 \label{fig:N113}
\end{figure*}

\begin{table*}
   \caption[]
           {Parameters of the Observed Molecular Lines toward the core of N113}
 \label{tbl:N113}
   \begin{flushleft}
   \begin{tabular}{l l c r c c r@{.}l@{$\pm$}l c}
   \hline
   \multicolumn{2}{l}{Molecule}      & \TMB  & r.m.s. & \vLSR
   & \delv  & \multicolumn{3}{c}{$\int$ \TMB\,d$v$} & Velocity Range \uspace1 \\
   \multicolumn{2}{l}{\& Transition} & [K]   & [mK]   & [\kms]
   & [\kms] & \multicolumn{3}{c}{[\Kkms]}          & [\kms] \dspace \\
   \hline
   \uspace2
   \HCC           & $N$=1--0 $J$=3/2--1/2 \bra1
            \begin{tabular}{l} $F$=2--1  \\ $F$=1--0  \end{tabular}
      & \hspace*{1ex} \begin{tabular}{c}  0.133    \\  0.075    \end{tabular}
                          &  27
      & \hspace*{1ex} \begin{tabular}{c}  235.7    \\  236.1    \end{tabular}
      & \hspace*{1ex} \begin{tabular}{c}  5.1      \\  4.8      \end{tabular}
      & \multicolumn{3}{c}{\hspace*{1ex} \begin{tabular}{r@{.}l@{$\pm$}l}
                                0&716 & 0.036 \\ 0&400 & 0.034 \end{tabular}}
      & \hspace*{1ex} \begin{tabular}{c} (230,242) \\ (230,241) \end{tabular} \\
   \uspace4
   HCN \see{a}    & $J$=1--0 \bra2
            \begin{tabular}{l} $F$=1--1  \\ $F$=2--1  \\ $F$=0--1  \end{tabular}
      & \hspace*{1ex} \begin{tabular}{c}  0.124    \\  0.202    \\
                                          0.083    \end{tabular}
                          &  46 & 234.6
      & \hspace*{1ex} \begin{tabular}{c}  7.2      \\  5.2      \\
                                          5.6      \end{tabular}
                                              &  2&44  & 0.10  & (220,252) \\
   \uspace1
   \HCOp          & $J$=1--0
                  & 0.583 &  53 & 235.1 & 5.4 &  3&29  & 0.07  & (229,241) \\
   \uspace1
   HNC            & $J$=1--0
                  & 0.144 &  40 & 234.7 & 5.8 &  0&867 & 0.055 & (228,241) \\
   \uspace1
   CS             & $J$=2--1
                  & 0.412 &  79 & 235.2 & 4.4 &  1&94  & 0.10  & (230,240) \\
   \uspace1
   \thCO          & $J$=1--0
                  & 1.28~~& 158 & 235.1 & 4.5 &  6&06  & 0.18  & (230,241) \\
   \uspace2
   CN \see{b}     & $N$=1--0 $J$=3/2--1/2 \bra1
            \begin{tabular}{l} $F$=3/2--1/2 \\ $F$=5/2--3/2 \end{tabular}
      & \hspace*{1ex} \begin{tabular}{c}   0.056   \\   0.172   \end{tabular}
                          &  56 & 235.4
      & \hspace*{1ex} \begin{tabular}{c}   5.7     \\   4.1     \end{tabular}
      & \multicolumn{3}{c}{\hspace*{1ex} \begin{tabular}{r@{.}l@{$\pm$}l}
                             0&319 & 0.060    \\ 0&753 & 0.053 \end{tabular}}
      & \hspace*{1ex} \begin{tabular}{c} (231,239) \\ (239,249) \end{tabular} \\
   \uspace1
   \twCO          & $J$=1--0
   \dspace        & 7.92~~& 570 & 235.3 & 5.2 & 44&1   & 0.7   & (230,242) \\
 \hline
   \end{tabular}
   \end{flushleft}
  {\footnotesize \begin{enumerate} \renewcommand{\labelenumi}{\alph{enumi})}
   \item The three HCN hyperfine transitions ($F$=1--1, $F$=2--1, $F$=0--1)
         have been resolved by a gaussian fit.
         While \TMB\ values for each component are given, the total integrated
         line intensity refers to the entire line.
   \item Only $J$=3/2--1/2 transitions of CN($N$=1--0)
         were covered by the high-resolution AOS backend; two of the
         hyperfine transitions ($F$=3/2--1/2 and $F$=5/2--3/2) were detected.
  \end{enumerate} }
\end{table*}

\begin{figure*}
   \vspace*{-1 cm}
   \hspace*{0 cm} \psfig{figure=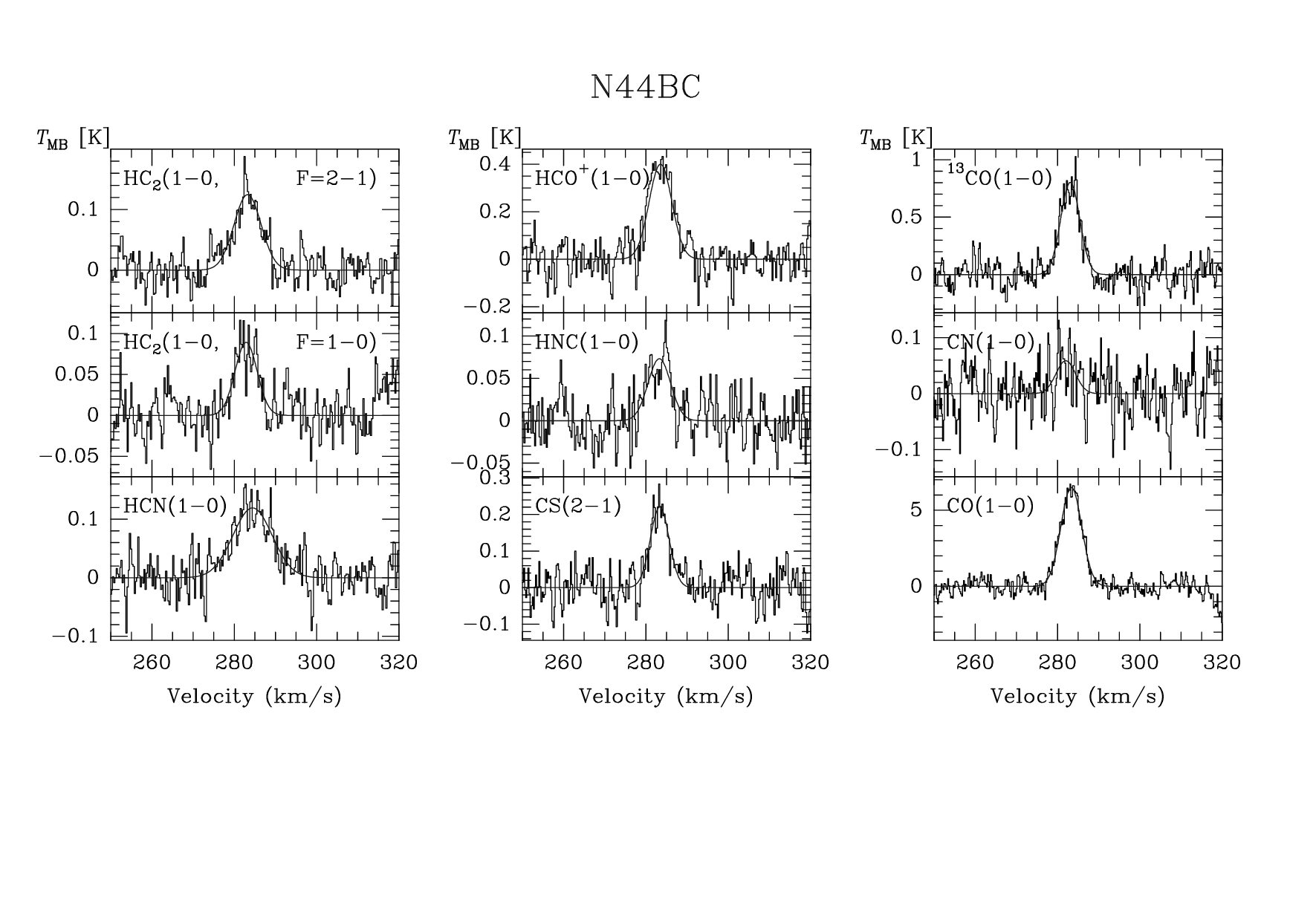,width=18 cm,angle=-90}
   \vspace*{-3 cm}
   \caption[]
           {Molecular spectra toward N44BC}
 \label{fig:N44BC}
\end{figure*}

\begin{table*}
   \caption[]
           {Parameters of the Observed Molecular Lines toward N44BC}
 \label{tbl:N44BC}
   \begin{flushleft}
   \begin{tabular}{l l c r c c r@{.}l@{$\pm$}l c}
   \hline
   \multicolumn{2}{l}{Molecule}      & \TMB  & r.m.s. & \vLSR
   & \delv  & \multicolumn{3}{c}{$\int$ \TMB\,d$v$} & Velocity Range \uspace1 \\
   \multicolumn{2}{l}{\& Transition} & [K]   & [mK]   & [\kms]
   & [\kms] & \multicolumn{3}{c}{[\Kkms]}          & [\kms] \dspace \\
   \hline
   \uspace2
   \HCC           & $N$=1--0 $J$=3/2--1/2 \bra1
            \begin{tabular}{l} $F$=2--1  \\ $F$=1--0  \end{tabular}
      & \hspace*{1ex} \begin{tabular}{c}  0.125    \\  0.089    \end{tabular}
                          &  26
      & \hspace*{1ex} \begin{tabular}{c}  283.2    \\  282.9    \end{tabular}
      & \hspace*{1ex} \begin{tabular}{c}  7.7      \\  6.3      \end{tabular}
      & \multicolumn{3}{c}{\hspace*{1ex}\begin{tabular}{r@{.}l@{$\pm$}l}
                           1&10  & 0.04  \\ 0&568 & 0.040 \end{tabular}}
      & \hspace*{1ex} \begin{tabular}{c} (274,292) \\ (275,291) \end{tabular} \\
   \uspace1
   HCN \see{a}    & $J$=1--0
                  & 0.119 &  32 & 284.4 &11.2~&  1&36  & 0.06  & (270,298) \\
   \uspace1
   \HCOp          & $J$=1--0
                  & 0.400 &  74 & 283.6 & 6.9 &  3&06  & 0.11  & (276,290) \\
   \uspace1
   HNC            & $J$=1--0
                  & 0.073 &  27 & 283.3 & 6.5 &  0&521 & 0.039 & (276,290) \\
   \uspace1
   CS             & $J$2--1
                  & 0.221 &  49 & 283.4 & 5.4 &  1&32  & 0.07  & (276,290) \\
   \uspace1
   \thCO          & $J$=1--0
                  & 0.802 & 112 & 283.1 & 5.5 &  4&70  & 0.14  & (276,290) \\
   \uspace1
   CN \see{b}     & $N$=1--0 $J$=3/2--1/2
                  & 0.059 &  47 & 282.0 & 5.6 &  0&378 & 0.059 & (276,290) \\
   \uspace1
   \twCO          & $J$=1--0
   \dspace        & 6.50~~& 475 & 283.3 & 6.0 & 40&2   & 0.6   & (276,290) \\
   \hline
   \end{tabular}
   \end{flushleft}
  {\footnotesize \begin{enumerate} \renewcommand{\labelenumi}{\alph{enumi})}
   \item The frequency of the main transition,
         HCN(1--0 $F$=2--1), is given.
         The three hyperfine transitions ($F$=1--1, $F$=2--1, $F$=0--1)
         remain unresolved.
         The relatively large spectral linewidth is
         caused by blending of these components.
   \item Only $J$=3/2--1/2 transitions of CN($N$=1--0)
         were covered by the high-resolution AOS backend;
         the $F$=5/2--3/2 hyperfine transition was detected.
  \end{enumerate} }
\end{table*}

\begin{figure*}
   \vspace*{-1 cm}
   \hspace*{0 cm} \psfig{figure=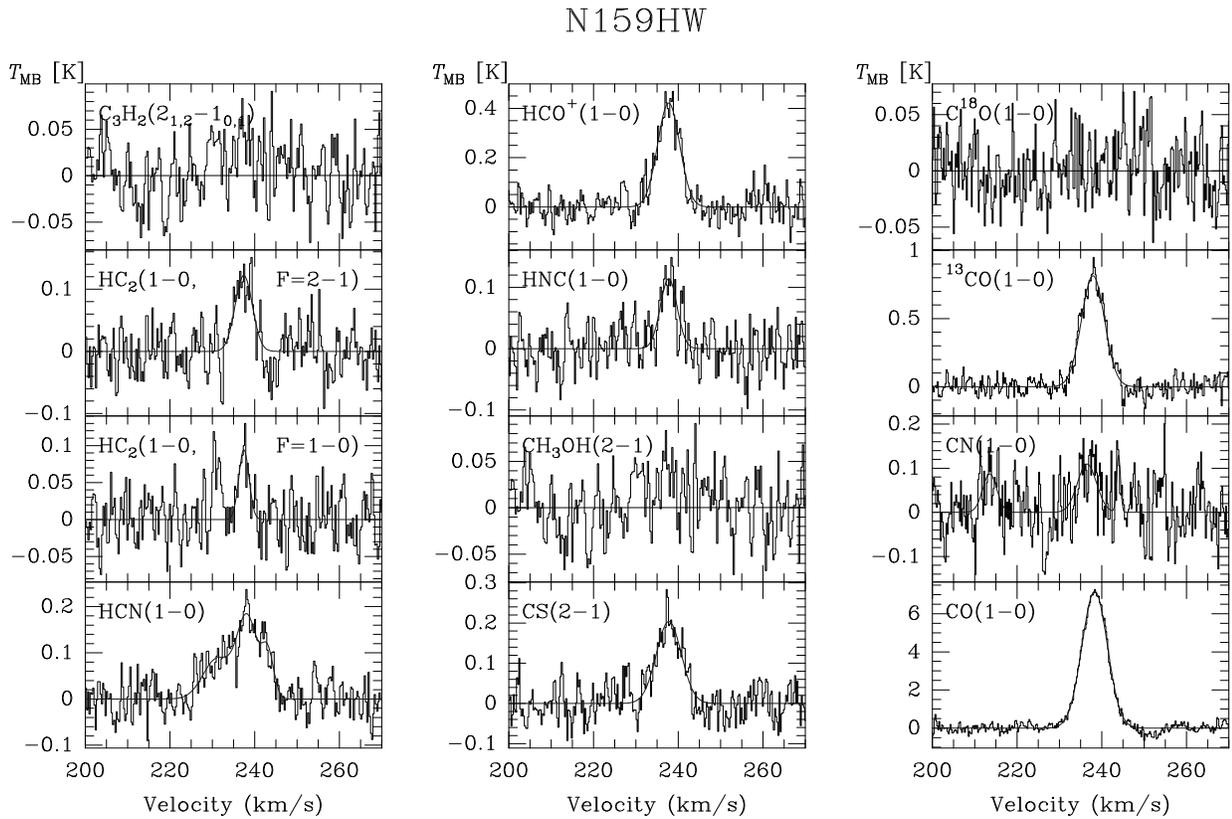,width=18 cm,angle=-90}
   \vspace*{-1 cm}
   \caption[]
           {Molecular spectra toward N159HW}
 \label{fig:N159HW}
\end{figure*}

\begin{table*}
   \caption[]
           {Parameters of the Observed Molecular Lines toward N159HW}
 \label{tbl:N159HW}
   \begin{flushleft}
   \begin{tabular}{l l c r c c r@{.}l@{$\pm$}l c}
   \hline
   \multicolumn{2}{l}{Molecule}      & \TMB  & r.m.s. & \vLSR
   & \delv  & \multicolumn{3}{c}{$\int$ \TMB\,d$v$} & Velocity Range \uspace1 \\
   \multicolumn{2}{l}{\& Transition} & [K]   & [mK]   & [\kms]
   & [\kms] & \multicolumn{3}{c}{[\Kkms]}          & [\kms] \dspace \\
   \hline
   \uspace1
   \CYCP          & $J$=2$_{1,2}-$1$_{0,1}$
          &     $<$ 0.05  &  32 & \ldots & \ldots
          & \multicolumn{3}{c}{$<$ 0.143 \see{c}}              & (230,245) \\
   \uspace2
   \HCC           & $N$=1--0 $J$=3/2--1/2 \bra1
            \begin{tabular}{l} $F$=2--1  \\ $F$=1--0  \end{tabular}
      & \hspace*{1ex} \begin{tabular}{c}  0.122    \\  0.094    \end{tabular}
                          &  38
      & \hspace*{1ex} \begin{tabular}{c}  237.4    \\  237.4    \end{tabular}
      & \hspace*{1ex} \begin{tabular}{c}  4.8      \\  2.8      \end{tabular}
      & \multicolumn{3}{c}{\hspace*{1ex} \begin{tabular}{r@{.}l@{$\pm$}l}
                                0&634 & 0.058 \\ 0&283 & 0.032 \end{tabular}}
      & \hspace*{1ex} \begin{tabular}{c} (230,245) \\ (235,240) \end{tabular} \\
   \uspace4
   HCN \see{a}    & $J$=1--0 \bra2
            \begin{tabular}{l} $F$=1--1  \\ $F$=2--1  \\ $F$=0--1  \end{tabular}
      & \hspace*{1ex} \begin{tabular}{c}  0.095    \\  0.175    \\
                                          0.087    \end{tabular}
                          &  39 & 238.2
      & \hspace*{1ex} \begin{tabular}{c}  3.4      \\  5.6      \\
                                          7.9      \end{tabular}
                                              &  2&16  & 0.08  & (224,251) \\
   \uspace1
   \HCOp          & $J$=1--0
                  & 0.422 &  55 & 237.7 & 6.1 &  2&93  & 0.08  & (230,245) \\
   \uspace1
   HNC            & $J$=1--0
                  & 0.114 &  38 & 237.7 & 4.7 &  0&551 & 0.057 & (230,245) \\
   \uspace1
   \METH          & $J$=2--1
          &     $<$ 0.05  &  32 & \ldots & \ldots
          & \multicolumn{3}{c}{$<$ 0.134 \see{c}}              & (230,245) \\
   \uspace1
   CS             & $J$=2--1
                  & 0.204 &  44 & 237.8 & 6.9 &  1&37  & 0.06  & (230,245) \\
   \uspace1
   \thCO          & $J$=1--0
                  & 0.814 &  61 & 237.9 & 6.4 &  5&42  & 0.08  & (230,245) \\
   \uspace4
   CN \see{b}     & $N$=1--0 $J$=3/2--1/2 \bra2
         \begin{tabular}{l}$F$=3/2--1/2\\$F$=5/2--3/2\\$F$=1/2--1/2\end{tabular}
      & \hspace*{1ex} \begin{tabular}{c}  0.048    \\  0.115    \\
                                          0.076    \end{tabular}
                          &  64 & 236.9
      & \hspace*{1ex} \begin{tabular}{c}  5.0      \\  4.6      \\
                                          3.6      \end{tabular}
      & \multicolumn{3}{c}{\hspace*{1ex} \begin{tabular}{r@{.}l@{$\pm$}l}
            0&267 & 0.053 \\ 0&512 & 0.060 \\ 0&241 & 0.053 \end{tabular}}
      & \hspace*{1ex} \begin{tabular}{c} (242,249) \\ (233,242) \\
                                         (210,217) \end{tabular} \\
   \uspace1
   \twCO          & $J$=1--0
   \dspace        & 7.17~~& 237 & 238.3 & 7.0 & 48&4   & 0.3   & (228,248) \\
   \hline
   \end{tabular}
   \end{flushleft}
  {\footnotesize \begin{enumerate} \renewcommand{\labelenumi}{\alph{enumi})}
   \item The three HCN hyperfine transitions ($F$=1--1, $F$=2--1, $F$=0--1)
         have been resolved by a gaussian fit.
         While \TMB\ values for each component are given, the total integrated
         line intensity refers to the entire line.
   \item Only $J$=3/2--1/2 transitions of CN($N$=1--0)
         were covered by the high-resolution AOS backend; three of the
         hyperfine transitions ($F$=3/2--1/2, $F$=5/2--3/2, and $F$=1/2--1/2)
         were detected.
   \item For undetected molecular lines, upper limits of 3 $\times$ r.m.s.
         are given for \hbox{$\int$\TMB\,d$v$}.
  \end{enumerate} }
\end{table*}

\begin{figure*}
   \vspace*{-1 cm}
   \hspace*{0 cm} \psfig{figure=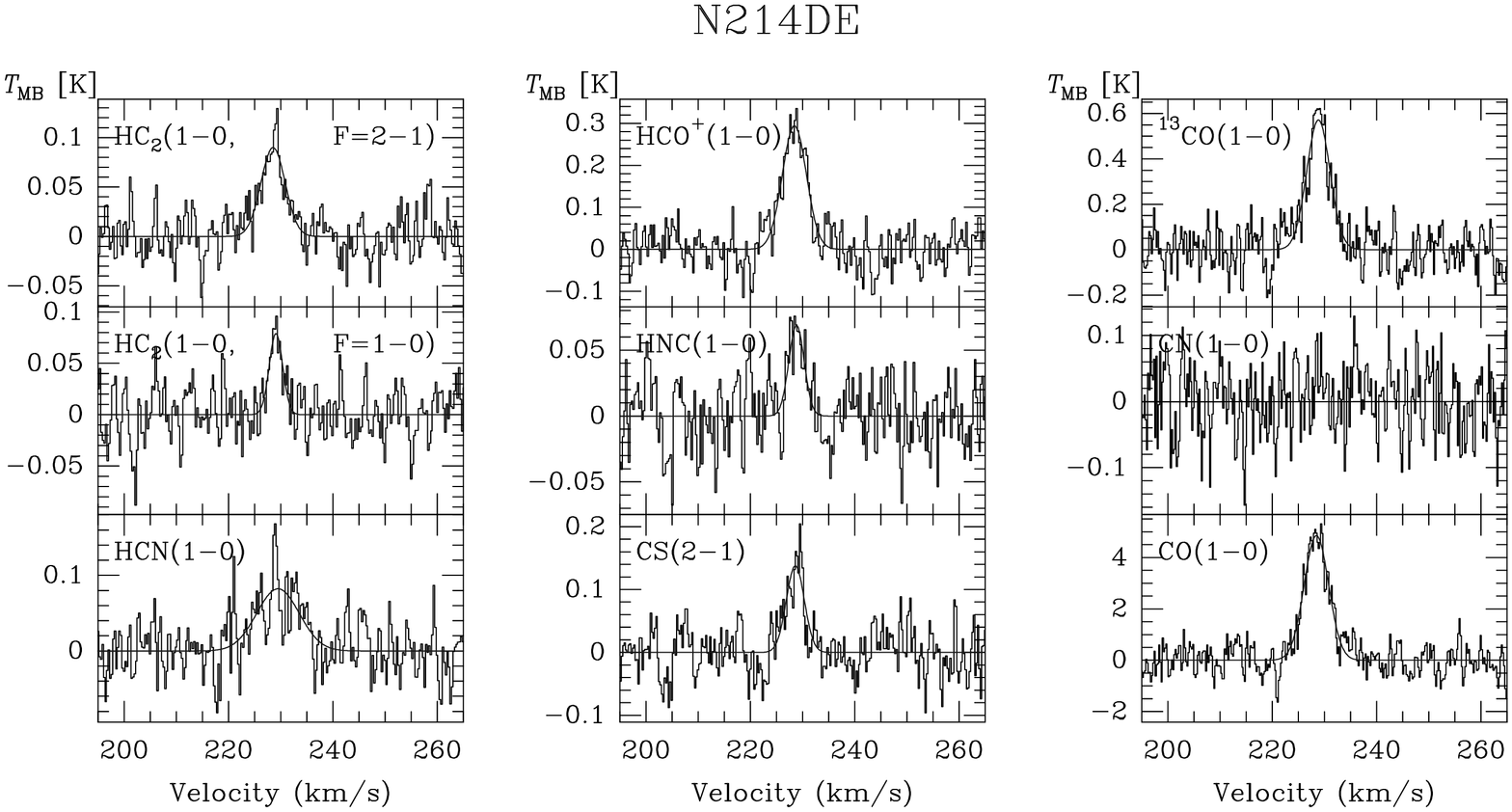,width=18 cm,angle=-90}
   \vspace*{-3 cm}
   \caption[]
           {Molecular spectra toward N214DE}
 \label{fig:N214DE}
\end{figure*}

\begin{table*}
   \caption[]
           {Parameters of the Observed Molecular Lines toward N214DE}
 \label{tbl:N214DE}
   \begin{flushleft}
   \begin{tabular}{l l c r c c r@{.}l@{$\pm$}l c}
   \hline
   \multicolumn{2}{l}{Molecule}      & \TMB  & r.m.s. & \vLSR
   & \delv  & \multicolumn{3}{c}{$\int$ \TMB\,d$v$} & Velocity Range \uspace1 \\
   \multicolumn{2}{l}{\& Transition} & [K]   & [mK]   & [\kms]
   & [\kms] & \multicolumn{3}{c}{[\Kkms]}          & [\kms] \dspace \\
   \hline
   \uspace2
   \HCC           & $N$=1--0 $J$=3/2--1/2 \bra1
            \begin{tabular}{l} $F$=2--1  \\ $F$=1--0  \end{tabular}
      & \hspace*{1ex} \begin{tabular}{c}  0.090    \\  0.079    \end{tabular}
                          &  23
      & \hspace*{1ex} \begin{tabular}{c}  228.5    \\  229.1    \end{tabular}
      & \hspace*{1ex} \begin{tabular}{c}  5.5      \\  3.0      \end{tabular}
      & \multicolumn{3}{c}{\hspace*{1ex} \begin{tabular}{r@{.}l@{$\pm$}l}
                                0&536 & 0.033 \\ 0&258 & 0.030 \end{tabular}}
      & \hspace*{1ex} \begin{tabular}{c} (222,235) \\ (224,235) \end{tabular} \\
   \uspace1
   HCN \see{a}    & $J$=1--0
                  & 0.082 &  33 & 229.5 & 9.6 &  0&831 & 0.055 & (220,239) \\
   \uspace1
   \HCOp          & $J$=1--0
                  & 0.294 &  47 & 228.4 & 5.4 &  1&75  & 0.06  & (222,235) \\
   \uspace1
   HNC            & $J$=1--0
                  & 0.069 &  27 & 228.8 & 3.5 &  0&278 & 0.037 & (222,235) \\
   \uspace1
   CS             & $J$=2--1
                  & 0.138 &  38 & 228.6 & 4.1 &  0&621 & 0.049 & (222,235) \\
   \uspace1
   \thCO          & $J$=1--0
                  & 0.571 &  92 & 228.8 & 5.3 &  3&36  & 0.11  & (222,235) \\
   \uspace1
   CN \see{b}     & $N$=1--0 $J$=3/2--1/2
          &     $<$ 0.06  &  59 & \ldots & \ldots
          & \multicolumn{3}{c}{$<$ 0.22  \see{c}}              & (222,235) \\
   \uspace1
   \twCO          & $J$=1--0
   \dspace        & 4.86~~& 566 & 228.5 & 5.5 & 29&0   & 0.7   & (222,235) \\
   \hline
   \end{tabular}
   \end{flushleft}
  {\footnotesize \begin{enumerate} \renewcommand{\labelenumi}{\alph{enumi})}
   \item The frequency of the main transition,
         HCN(1--0 $F$=2--1), is given.
         The three hyperfine transitions ($F$=1--1, $F$=2--1, $F$=0--1)
         remain unresolved.
         The relatively large spectral linewidth is
         caused by blending of these components.
   \item Only $J$=3/2--1/2 transitions of CN($N$=1--0)
         were covered by the high-resolution AOS backend;
         the $F$=5/2--3/2 hyperfine transition was detected.
   \item For undetected molecular lines, upper limits of 3 $\times$ r.m.s.
         are given for \hbox{$\int$\TMB\,d$v$}.
  \end{enumerate} }
\end{table*}

\begin{figure*}
   \vspace*{-1 cm}
   \hspace*{0 cm} \psfig{figure=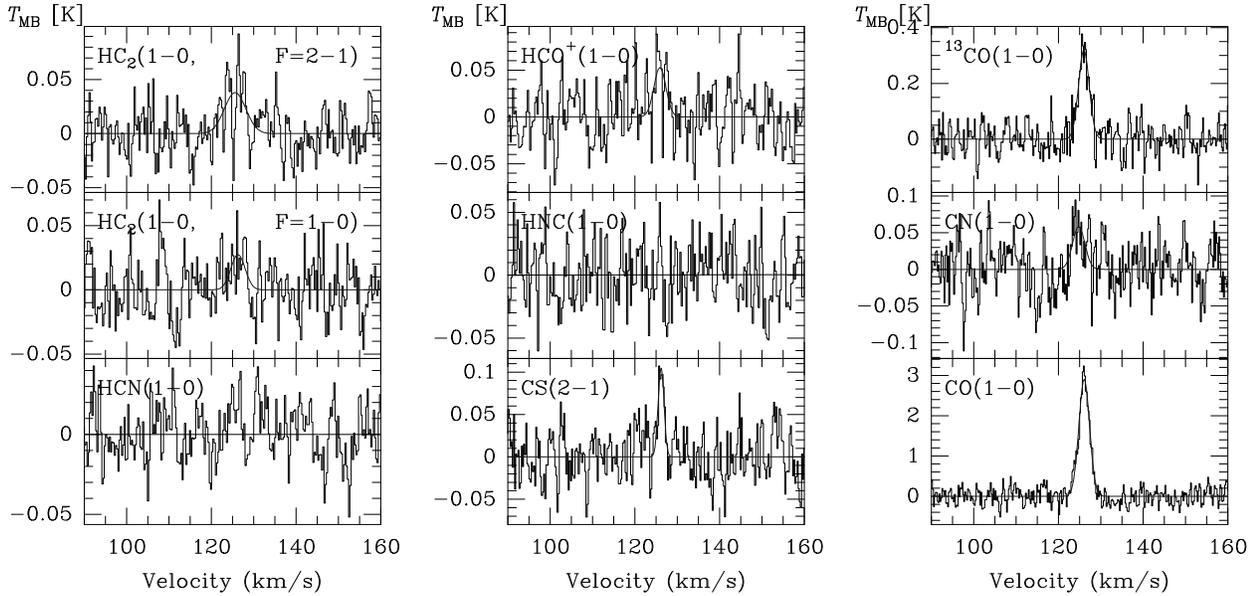,width=18 cm,angle=-90}
   \vspace*{-3 cm}
   \caption[]
           {Molecular spectra toward LIRS\,36}
 \label{fig:LIRS36}
\end{figure*}

\begin{table*}
   \caption[]
           {Parameters of the Observed Molecular Lines toward LIRS\,36}
 \label{tbl:LIRS36}
   \begin{flushleft}
   \begin{tabular}{l l c r c c r@{$\pm$}l c}
   \hline
   \multicolumn{2}{l}{Molecule}      & \TMB  & r.m.s. & \vLSR
   & \delv  & \multicolumn{2}{c}{$\int$ \TMB\,d$v$} & Velocity Range \uspace1 \\
   \multicolumn{2}{l}{\& Transition} & [K]   & [mK]   & [\kms]
   & [\kms] & \multicolumn{2}{c}{[\Kkms]}          & [\kms] \dspace \\
   \hline
   \uspace2
   \HCC           & $N$=1--0 $J$=3/2--1/2 \bra1
            \begin{tabular}{l} $F$=2--1  \\ $F$=1--0  \end{tabular}
      & \hspace*{1ex} \begin{tabular}{c}  0.038    \\  0.027    \end{tabular}
                          &  21
      & \hspace*{1ex} \begin{tabular}{c}  125.6    \\  126.3    \end{tabular}
      & \hspace*{1ex} \begin{tabular}{c}  6.0      \\  4.0      \end{tabular}
      & \multicolumn{2}{c}{\begin{tabular}{r@{$\pm$}l}
                                0.223 & 0.027 \\ 0.093 & 0.024 \end{tabular}}
      & \hspace*{1ex} \begin{tabular}{c} (120,131) \\ (122,131) \end{tabular} \\
   \uspace1
   HCN            & $J$=1--0
          &     $<$ 0.02  &  18 & \ldots& \ldots
          & \multicolumn{2}{c}{$<$ 0.064 \see{b}}              & (121,131) \\
   \uspace1
   \HCOp          & $J$=1--0
                  & 0.053 &  30 & 125.9 & 3.3 &  0.229 & 0.036 & (121,131) \\
   \uspace1
   HNC            & $J$=1--0
          &     $<$ 0.02  &  26 & \ldots & \ldots
          & \multicolumn{2}{c}{$<$ 0.092 \see{b}}              & (121,131) \\
   \uspace1
   CS             & $J$=2--1
                  & 0.105 &  30 & 126.2 & 1.6 &  0.276 & 0.035 & (121,131) \\
   \uspace1
   \thCO          & $J$=1--0
                  & 0.319 &  57 & 126.0 & 3.0 &  0.923 & 0.061 & (121,131) \\
   \uspace1
   CN \see{a}     & $N$=1--0 $J$=3/2--1/2
                  & 0.058 &  34 & 124.8 & 3.4 &  0.154 & 0.037 & (121,131) \\
   \uspace1
   \twCO          & $J$=1--0
   \dspace        & 2.90  & 213 & 126.1 & 3.0 &  9.26  & 0.23  & (121,131) \\
   \hline
   \end{tabular}
   \end{flushleft}
  {\footnotesize \begin{enumerate} \renewcommand{\labelenumi}{\alph{enumi})}
   \item Only $J$=3/2--1/2 transitions of CN($N$=1--0)
         were covered by the high-resolution AOS backend;
         the $F$=5/2--3/2 hyperfine transition was detected.
   \item For undetected molecular lines, upper limits of 3 $\times$ r.m.s.
         are given for \hbox{$\int$\TMB\,d$v$}.
  \end{enumerate} }
\end{table*}

\begin{figure*}
   \vspace*{-1 cm}
   \psfig{figure=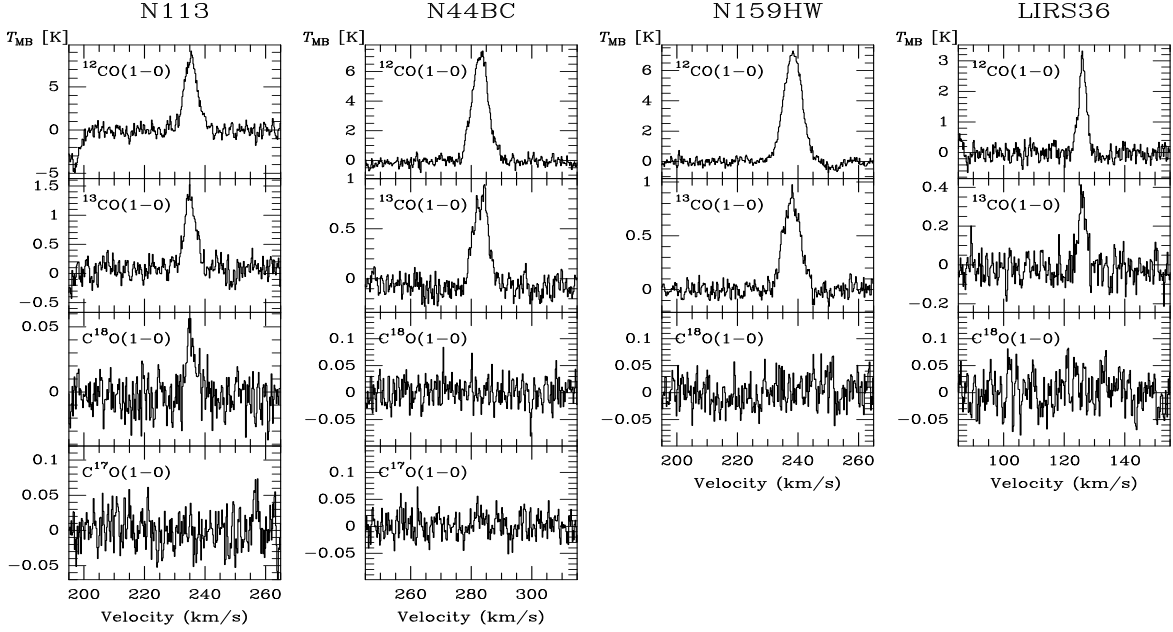,width=18 cm,height=10 cm,angle=-90}
   \vspace*{-1 cm}
   \caption[]
           {The spectra of the observed carbon monoxide isotopomers CO, \thCO,
            \CeiO, and \CseO\ toward {\bf a)} N113, {\bf b)} N44BC,
            {\bf c)} N159HW, and {\bf d)} LIRS\,36.}
 \label{fig:MC1-isotope}
\end{figure*}

\begin{table*}
   \caption[]
           {Line parameters of carbon monoxide isotopomers}
 \label{tbl:MC1-isotope}
   \begin{flushleft}
   \begin{tabular}{l c r@{.}l r c c r@{.}l@{$\pm$}l c}
   \hline
   \uspace1
   Object      &  Molecule & \multicolumn{2}{c}{\TMB}   & r.m.s.  &  \vLSR
               & \delv     & \multicolumn{3}{c}{$\int$ \TMB\,d$v$}
               & Velocity Range \\
               &           & \multicolumn{2}{c}{[K]}     &  [mK]   &  [\kms]
               & [\kms]    & \multicolumn{3}{c}{[\Kkms]}
   \dspace     & [\kms]         \\
   \hline
   \uspace1
 N113          & \twCO   &     7&92  &   570 &   235.3 &  5.2
                                           & 32&4   & 0.3   &  (228,242) \\
               & \thCO   &     1&28  &   158 &   235.1 &  4.5
                                           &  4&28  & 0.07  &  (228,242) \\
               & \CeiO   &     0&439 &    12 &   235.3 &  3.3
                                           &  0&169 & 0.014 &  (228,242) \\
               & \CseO   & $<$ 0&02  &    25 &  \ldots &  \ldots
               & \multicolumn{3}{c}{$<$ 0.096 \see{a}}      &  (228,242) \\
\\
 N44BC         & \twCO   &     6&55  &   475 &   283.3 &  6.0
                                           & 40&2   & 0.6   &  (276,290) \\
               & \thCO   &     0&802 &   112 &   283.1 &  5.5
                                           &  4&70  & 0.14  &  (276,290) \\
               & \CeiO   & $<$ 0&02  &    25 &  \ldots &  \ldots
               & \multicolumn{3}{c}{$<$ 0.092 \see{a}}      &  (276,290) \\
               & \CseO   & $<$ 0&02  &    22 &  \ldots &  \ldots
               & \multicolumn{3}{c}{$<$ 0.085 \see{a}}      &  (276,290) \\
\\
 N159HW        & \twCO   &     7&17  &   237 &   238.3 &  7.0
                                           & 52&4   & 0.4   &  (228,248) \\
               & \thCO   &     0&814 &    61 &   237.9 &  6.4
                                           &  5&75  & 0.08  &  (230,245) \\
               & \CeiO   & $<$ 0&03  &    29 &  \ldots &  \ldots
               & \multicolumn{3}{c}{$<$ 0.116 \see{a}}      &  (230,245) \\
\\
 LIRS\,36      & \twCO   &     2&90  &   213 &   126.1 &  3.0
                                           &  9&26  & 0.23  &  (121,131) \\
               & \thCO   &     0&319 &    57 &   126.0 &  3.0
                                           &  0&923 & 0.061 &  (121,131) \\
               & \CeiO   & $<$ 0&02  &    34 &  \ldots &  \ldots
   \dspace     & \multicolumn{3}{c}{$<$ 0.110 \see{a}}      &  (121,131) \\
   \hline
   \end{tabular}
   \end{flushleft}
  {\footnotesize \begin{enumerate} \renewcommand{\labelenumi}{\alph{enumi})}
   \item For undetected CO isotopomers, upper 3\,$\sigma$ limits
         are given for \hbox{$\int$\TMB\,d$v$}.
  \end{enumerate} }
\end{table*}

\section{Results}
\label{sec:MC1-Results}

\subsection{CO maps toward HII regions in the LMC}
\label{sec:MC1-maps}

   $^{13}$CO maps defining the molecular clouds were obtained with a
   spacing of 30\arcsec\ toward the \HII\ regions N79, N83A, N11, N105A,
   N113, N44BC, N55A, N160, and N214DE (20\arcsec\ spacing) in the LMC.
   In addition, \twCO\ maps with the same spacing were obtained
   for N113, N44BC, and N214DE (see Table\,\ref{tbl:MC1-source}).
   The \thCO(1--0) transition is very effective for mapping molecular clouds
   in the Magellanic Clouds, because the emission providing direct information
   on column density is strong enough for short integration times.
   The contour maps are shown in Fig.\,\ref{fig:contour},
   Fig.\,\ref{fig:contour-N113}.b, Fig.\,\ref{fig:contour-N44BC}.b,
   and Fig.\,\ref{fig:contour-N214DE}.b.
   In most cases, the offsets of the \thCO\ peak
   from the \Halpha\ center are within the pointing error of the SEST.
   Nevertheless, this is not the case for N113, N214DE, and N160,
   where right ascension and declination offsets are $\approx$
   ($-$10\arcsec, $-$20\arcsec), $\approx$ ($-$15\arcsec, 20\arcsec),
   and $\approx$ ($-$35\arcsec, $-$20\arcsec), respectively.
   In N113 the position of the \thCO\ peak is confirmed by a \twCO\ map
   obtained with a spacing of 20\arcsec\ in the frequency-switching mode.
   The map shows a compact core which is similar
   to that observed in \thCO\ (Fig.\,\ref{fig:contour-N113}.a).
   However, contour maps of N44BC and N214DE
   (Fig.\,\ref{fig:contour-N44BC} and Fig.\,\ref{fig:contour-N214DE})
   show quite different \twCO\ and \thCO\ distributions.

\subsection{Multiline study of the Magellanic Clouds}
\label{sec:MC1-multiline}

   Previously, a comprehensive molecular-line study of an individual Magellanic
   Cloud \HII\ region has been made only for N159 (Johansson \etal\ 1994).
   While interesting differences between the interstellar medium
   of the LMC and that of the Galaxy were found,
   results for one object may not be typical for the entire galaxy.
   Furthermore, differences in the properties of the rarer molecular species
   between the LMC and the SMC have not been studied.
   To further our understanding in these areas, we have observed a variety of
   $\lambda$=3\,mm molecular transitions toward the molecular cores
   of N113, N44BC, N159HW, and N214DE in the LMC, and LIRS\,36 in the SMC.
   The spectra are shown in Figs.\,\ref{fig:N113} to \ref{fig:LIRS36},
   respectively.
   The corresponding line parameters of the molecular species,
   including \HCC, HCN, \HCOp, HNC, CS, \thCO, CN, and \twCO, are summarized
   in Tables\,\ref{tbl:N113} -- \ref{tbl:LIRS36} (for the line frequencies
   and antenna beamwidths, see also Table\,\ref{tbl:MC1-freq}).
   Toward some of the sources, individual hyperfine components
   of HCN and CN can be identified.
   While relative intensities of the CN features are consistent with Local
   Thermodynamic Equilibrium (LTE) and optically thin line emission,
   the HCN features show deviations from LTE
   which are discussed in Sect.\ref{sec:MC1-hnc}.

\subsection{Observations of CO isotopomers in the Magellanic Clouds}
\label{sec:MC1-isotope}

   To gain insight into the isotopic composition of the interstellar medium of
   the Magellanic Clouds, three carbon monoxide isotopomers
   (\twCO, \thCO, and \CeiO) were observed toward the
   molecular cores of N113, N44BC, N159HW, and LIRS\,36,
   and \CseO\ was also measured toward N113 and N44BC.
   The spectra are shown in Fig\,\ref{fig:MC1-isotope}.
   Table\,\ref{tbl:MC1-isotope} displays the corresponding line parameters.
   Only upper limits could be obtained in \CeiO\ and \CseO\
   (see also Sect.\,\ref{sec:MC1-c18o}).

\section{Discussion}
\label{sec:MC1-Discussion}

   In the following, cloud stability, $N_{\rm H_2}/I_{\rm CO}$ conversion
   factors, cloud chemistry, and isotope ratios will be discussed
   for the molecular cores we have observed near prominent \HII\ regions.

\begin{table*}
   \caption[]
           {Cloud parameters and gravitational stability}
 \label{tbl:MC1-stable}
   \begin{flushleft}
   \begin{tabular}{l r@{}l r@{}l c c r@{}l r@{.}l c c r}
   \hline
   \uspace1
   Source  & \multicolumn{2}{c}{$\alpha_{\rm obs}$}
           & \multicolumn{2}{c}{$\alpha$}  & $D$ \see{a}    & \delv
           & \multicolumn{2}{c}{\Mvir}
           & \multicolumn{2}{c}{\Eturb}
           & \Etherm                       & \Egrav         & $B_{\rm equi}$ \\
   \dspace & \multicolumn{2}{c}{[\arcsec]}
           & \multicolumn{2}{c}{[\arcsec]} & [pc]           & [\kms]
           & \multicolumn{2}{c}{[10$^3$ \Msol]}
           & \multicolumn{2}{c}{[10$^{42}$ J]}
           & [10$^{39}$ J]                 & [10$^{42}$ J]  & [$\mu$G]       \\
   \hline
   \uspace1
   N79    &  90 &~&  78 &~&  19  &  5.3  &   64 &&~~0&97 &  24  &  2.2  &  55 \\
   N83A   &  75 & &  60 & &  15  &  6.0  &   63 &&  1&2  &  23  &  2.8  &  92 \\
   N11    &  80 & &  66 & &  16  &  5.3  &   55 &&  0&82 &  20  &  1.9  &  65 \\
   N105A  &  80 & &  66 & &  16  &  7.2  &~~101 &&  2&8  &  37  &  6.5  & 120 \\
   N113   & 100 & &  89 & &  22  &  4.5  &   53 &&  0&58 &  20  &  1.3  &  34 \\
   N44BC  & 120 & & 111 & &  27  &  5.6  &  102 &&  1&7  &  38  &  4.0  &  43 \\
   N55A   &  75 & &  60 & &  15  &  6.4  &   72 &&  1&6  &  27  &  3.7  & 104 \\
   N160   &  95 & &  84 & &  20  &  7.3  &  131 &&  3&7  &  48  &  8.7  &  98 \\
   N214DE & 110 & & 100 & &  24  &  4.7  &   65 &&  0&77 &  24  &  1.8  &  34 \\
   \hline
   \end{tabular}
   \end{flushleft}
   {\footnotesize \begin{enumerate} \renewcommand{\labelenumi}{\alph{enumi})}
                                                  \renewcommand{\itemsep}{0pt}
   \item The assumed distance to the LMC is 50\,kpc.
   \end{enumerate} }
\end{table*}

\subsection{Gravitational stability of the mapped clouds}
\label{sec:MC1-stable}

   For the molecular clouds which have been mapped in the \thCO(1--0)
   transition, we can estimate the virial mass, which depends only on
   the linewidth \delv\ and the (intrinsic) cloud diameter $D$.
   Assuming a spherical cloud geometry and constant gas density in the clouds,
   \Mvir\ can be obtained from
\begin{equation}
     M_{\rm vir} = \frac{5}{8\,\sqrt{3}\,G\,\ln 2}\,D\,(\Delta v_{1/2})^2\ ,
\end{equation}
   where $G$ is the gravitational constant.
   In terms of astronomically convenient units, this relation becomes
\begin{equation}
     M_{\rm vir}\,[{\rm M}_\odot] = 121\ D\,[{\rm pc}]\
                                \Delta v_{1/2}\,[{\rm km\,s}^{-1}]\ ,
 \label{eqn:Mvir}
\end{equation}
   which is correct to within 10\% also for a $1/r$ density profile, and
   to a factor of two if $n$(\MOLH) $\propto$ $1/r^2$
   (\eg\ MacLaren \etal\ 1988).
   The angular cloud diameter $\alpha$ can be calculated from the observed
   angular diameter $\alpha_{\rm obs}$ defined by the contour maps with
\begin{equation}
     \alpha^2 = \alpha_{\rm obs}^2 - HPBW^2 \ ,
 \label{eqn:diameter}
\end{equation}
   where $HPBW$, the half power beam width, $\approx$ 45\arcsec\ at
   the frequency of the \thCO\ $J$=1--0 transition.

   The stability of the clouds can be evaluated by comparing
   the turbulent kinetic, thermal, and gravitational energy
   (\eg\ Harju \etal\ 1992).
   The turbulent kinetic energy \Eturb\ can be estimated from
\begin{equation}
     E_{\rm turb} = \frac{3}{2} M_{\rm vir} \sigma ^2 \ .
 \label{eqn:Eturb}
\end{equation}
   The one-dimensional turbulent velocity dispersion $\sigma$ is
   related to the (intrinsic) linewidth by
\begin{equation}
     \sigma^2 = \frac{\Delta v_{\rm int}^2}{8\ln2} -
                \frac{kT}{m_{^{13}{\rm CO}}}\ ,
\end{equation}
   where $m_{^{13}{\rm CO}}$ is the mass of the \thCO\ molecule;
   $k$ is the Boltzmann constant, and $T$ is the kinetic temperature,
   assumed to be 30\,K (see \eg\ Figs.\,2a and 5a of Lequeux \etal\ 1994).
   The intrinsic line width $\Delta v_{\rm int}$ is derived from the observed
   line width $\Delta v_{\rm obs}$ and the velocity resolution
   $\Delta v_{\rm res}$ (0.12\,\kms\ for our \thCO(1--0) measurement) by:
\begin{equation}
     \Delta v_{\rm int}^2 = \Delta v_{\rm obs}^2 - \Delta v_{\rm res}^2\ ;
\end{equation}
   or $\Delta v_{\rm int} \approx \Delta v_{\rm obs}$,
   if $\Delta v_{\rm obs} \gg \Delta v_{\rm res}$.

   The thermal energy can be estimated from
\begin{equation}
     E_{\rm therm} = \frac{3}{2} NkT\ ,
 \label{eqn:Etherm}
\end{equation}
   where $N$ is the total number of \MOLH\ molecules.
   An upper limit for $N$ can be derived from $N$ = \Mvir/$m_{{\rm H}_2}$.

   Assuming a homogeneous sphere, the gravitational energy can be expressed as:
\begin{equation}
     E_{\rm grav} = \frac{6}{5}\,G\,\frac{M_{\rm vir}^2}{D}\ .
 \label{eqn:Egrav}
\end{equation}

   The cloud parameters of most of the mapped \HII\ regions are displayed
   in Table\,\ref{tbl:MC1-stable}.
   The gravitational energy is larger than the turbulent kinetic energy
   and much larger than the upper limit of thermal energy,
   \ie, \Egrav\ $>$ \Eturb\ $\gg$ \Etherm.
   Thus all listed \HII\ regions seem to be gravitationally unstable.
   Taking the magnetic energy density $B^2/8\pi$ and assuming a uniform
   magnetic field strength, the magnetic fields required
   for cloud equilibrium are also given in Table\,\ref{tbl:MC1-stable}.
   While the results can only be judged to be order of magnitude estimates,
   it is interesting to see that the calculated magnetic field strengths
   of several 10 to 120\,$\mu$G are slightly larger than the
   observed magnetic fields in star-forming regions which only refer to
   the large scale magnetic component parallel to the line of sight
   (for a summary, see \eg\ Fiebig \& G\"usten 1989; Heiles \etal\ 1993;
   Vall\'ee 1995).

   For the observed \HII\ regions the virial masses have been obtained
   and the gravitational stability has been checked
   (see Sect.\,\ref{sec:MC1-stable} and Table\,\ref{tbl:MC1-stable}).
   Since the \twCO(1--0) emission has also been measured toward N113 and N44BC,
   the \MOLH\ mass-to-CO luminosity conversion factor can be derived with
\begin{equation}
     M_{\rm gas} = 1.36\,[N_{\rm H_2}/I_{\rm CO}]\,m_{\rm H_2}\,
                   \sum_{i = 1}^n \frac{I_{u,i} + I_{l,i}}{2}\,A_i\ ,
 \label{eqn:co2h2}
\end{equation}
   where 1.36 is the correction to include helium and metals (likely an
   overestimate but consistent with the value taken in other studies),
   $m_{\rm H_2}$ is the mass of an \MOLH\ molecule,
   $I_{u,i}$ and $I_{l,i}$ are the integrated intensities of
   the upper and lower contours confining the area $A_i$
   (in cm$^{-2}$; $n$: number of contours).
   $X = [N_{\rm H_2}/I_{\rm CO}]$ is the conversion factor
   between integrated \twCO\ $J$=1--0 intensity (in \Kkms) and
   \MOLH\ column density (in cm$^{-2}$).
   For our Galactic disk $X = 2.3\times 10^{20}$\,mol\,cm$^{-2}$\,
   (\Kkms)$^{-1}$ is standard (Strong \etal\ 1988).
   This conversion factor was based on the CO survey by Dame \etal\ (1987).
   A re-evaluation of the calibration scheme at the Columbia telescopes
   (Bronfman \etal\ 1988) has shown that the CO intensity scale was
   too low by 22\%, thus yielding a conversion factor
   $X_{\rm Gal} = 1.9\times 10^{20}$\,mol\,cm$^{-2}$\,(\Kkms)$^{-1}$.
   In the LMC $X_{\rm LMC} \approx 6\,X_{\rm Gal}$
   was determined by Cohen \etal\ (1988).
   Assuming \Mgas\ = \Mvir\ we get conversion factors of
   $X_{\rm N113} = 1.8\times 10^{20}$\,mol\,cm$^{-2}$\,(\Kkms)$^{-1}$ and
   $X_{\rm N44BC} = 2.4\times 10^{20}$\,mol\,cm$^{-2}$\,(\Kkms)$^{-1}$,
   which are smaller than the value from Cohen \etal\ (1988),
   but close to that in the Galactic disk and that determined
   by Garay \etal\ (1993) for the 30\,Doradus halo where a value of
   $2.8\times 10^{20}$\,mol\,cm$^{-2}$\,(\Kkms)$^{-1}$ was determined.
   For the SMC Rubio \etal\ (1993b) found with $X_{\rm SMC} = 9\times 10^{20}$\,
   ($R$/10\,pc)$^{0.7}$\,mol\,cm$^{-2}$\,(\Kkms)$^{-1}$
   a dependence on linear scale length (\ie, beam radius).
   It seems that the large \MOLH\ mass-to-CO luminosity conversion factor
   determined by Cohen \etal\ (1988) for the LMC is caused by a similar
   scale length dependence.
   This can be explained by the fact that the molecular clouds are, in general,
   smaller than the beam size and that the interclump gas is mainly
   in an atomic phase, not contributing to the integrated CO emission.
   For compact molecular hot spots in the LMC, the conversion factor
   appears to be close to the Galactic disk value,
   in good agreement with the result of Johansson (1991).

\subsection{Molecular intensity ratios}
\label{sec:MC1-ratio}

\begin{table*}
   \caption[]
           {Integrated $J$=1--0 or $N$=1--0 intensity ratios in the observed
            \HII\ regions, in Galactic molecular clouds and
            in some external galaxies.
            For IC\,443 and W49N, we give abundance ratios. \see{a}}
 \label{tbl:MC1-ratio}
   \begin{flushleft}
   \begin{tabular}{l r@{.}l r@{.}l r@{.}l r@{.}l r@{.}l r@{.}l r@{}l}
   \hline
   \uspace2 \dspace
   Sources         & \multicolumn{2}{c}{$\frac{\HCOp}{\hbox{HCN}}$}
                   & \multicolumn{2}{c}{$\frac{\hbox{HCN}}{\hbox{HNC}}$}
                   & \multicolumn{2}{c}{$\frac{\hbox{CN}}{\hbox{HCN}}$}
                   & \multicolumn{2}{c}{$\frac{\HCC}{\hbox{HCN}}$}
                   & \multicolumn{2}{c}{$\frac{\hbox{CS}}{\hbox{CN}}$}
                   & \multicolumn{2}{c}{$\frac{\twCO}{\thCO}$}
                   & \multicolumn{2}{c}{$\frac{\thCO}{\CeiO}$} \\
   \hline
   \uspace1
   N113          &    1&35 & ~~~2&82 &    0&44 &    0&46
                           &    1&81 & ~~~7&28 &    36&   \\
   N44BC         &    2&25 &    2&61 &    0&28 &    1&22
                           &    3&50 &    8&56 & $>$48&   \\
   N159HW        &    1&36 &    3&90 &    0&52 &    0&42
                           &    1&26 &    9&12 & $>$49&   \\
   N214DE        &    2&10 &    2&99 & $<$0&26 &    0&96
                           & $>$2&89 &    8&64 & \multicolumn{2}{c}{\ldots} \\
   LIRS\,36      & $>$3&59 & \multicolumn{2}{c}{\ldots}  & $>$2&60 & $>$5&5
                           &    1&66 &   10&0  &  $>$8&   \\
 \uspace1
   N159 \see{b}  &    2&21 &    2&97 &    0&48 & \multicolumn{2}{c}{\ldots}
                           &    1&75 &    8&31 &    34&   \\
 \uspace1
   S138 \see{b}  &    0&67 &    3&75 & \multicolumn{2}{c}{\ldots}
                 & \multicolumn{2}{c}{\ldots}  & \multicolumn{2}{c}{\ldots}
                                     &    5&59 &    10&   \\
   M17SW \see{c} &    0&16 &    1&9  & \multicolumn{2}{c}{$\ll$ 1}
                 & \multicolumn{2}{c}{\ldots}  & \multicolumn{2}{c}{\ldots}
                                     &    4&9  &     8&.3 \\
   IRC+10216 \see{d}
                 & \multicolumn{2}{c}{\ldots}  &    8&24 &    1&24
                 & \multicolumn{2}{c}{\ldots}  &    0&36 &   10&8
                 & \multicolumn{2}{c}{\ldots} \\
   IC\,443 \see{e}
                 &    1&2  &    7&0  &    1&7  & \multicolumn{2}{c}{\ldots}
                           &    0&5  & \multicolumn{2}{c}{\ldots}
                                     & \multicolumn{2}{c}{\ldots} \\
   W49N \see{f}
                 &    1&0  &    6&0  &    0&5  &$>$25&0  &    2&0
                 & \multicolumn{2}{c}{\ldots}  & \multicolumn{2}{c}{\ldots} \\
 \uspace1
   NGC\,253      & 0&81 \see{g} & 1&3  \see{h} & 1&5  \see{g} & 0&20 \see{i}
                 & 0&51 \see{j} & 16&6 \see{k} & 4&.9 \see{l} \\
   IC\,342       & 0&51 \see{m} & 3&3  \see{h} & 0&79 \see{n}
                                               & \multicolumn{2}{c}{\ldots}
                 & 0&32 \see{o} & 11&1 \see{k} & 3&.9 \see{l} \\
   M\,82         & 2&1  \see{p} & 2&0  \see{h} & 1&4  \see{n} & 0&46 \see{q}
                 & 0&68 \see{o} & 15&9 \see{k} & 4&.5 \see{l} \\
   NGC\,4945     & 0&93 \see{r} & 2&03 \see{r} & 1&48 \see{s} & 0&45 \see{r}
                 & 0&35 \see{r} & 14&4 \see{s} & 2&.9 \see{s} \\
   \hline
   \end{tabular}
   \end{flushleft}
   {\footnotesize \begin{enumerate} \renewcommand{\labelenumi}{\alph{enumi})}
                  \setlength{\baselineskip}{10pt} \renewcommand{\itemsep}{0pt}
   \item Transitions are $J$=1--0 or $N$=1--0 for all molecular species
         except for CS.
         For CS, $J$=2--1 intensities were taken.
         CN line intensities refer to detected $N$=1--0 hyperfine components.
   \item Johansson \etal\ (1994).
   \item Baudry \etal\ (1980), Lada (1976), and Turner \& Thaddeus (1977).
   \item Nyman \etal\ (1993).
   \item Ziurys \etal\ (1989).
   \item Nyman \& Millar (1989).
   \item Henkel \etal\ (1993).
   \item H\"uttemeister \etal\ (1995).
   \item Israel (1992).
   \item CS data from Mauersberger \& Henkel (1989),
         CN data from Henkel \etal\ (1993).
   \item Sage \& Isbell (1991).
   \item Sage \etal\ (1991).
   \item Nguyen-Q-Rieu \etal\ (1992).
   \item CN data from Henkel \etal\ (1988),
         HCN data from Nguyen-Q-Rieu \etal\ (1992).
   \item CS data from Mauersberger \& Henkel (1989),
         CN data from Henkel \etal\ (1988).
   \item Nguyen-Q-Rieu \etal\ (1989).
   \item HC$_2$ data from Henkel \etal\ (1988),
         HCN data from Nguyen-Q-Rieu \etal\ (1989).
   \item Henkel \etal\ (1990).
   \item Henkel \etal\ (1994).
   \end{enumerate} }
\end{table*}

   In Table\,\ref{tbl:MC1-ratio} we present the observed integrated
   line intensity ratios for the LMC and the SMC.
   A comparison of these ratios with those from
   Johansson \etal\ (1994) for N159, Galactic \HII\ regions, the shocked
   molecular gas associated with the supernova remnant, IC\,443,
   the diffuse absorbing gas toward W49N, and a sample of
   prominent nearby galaxies is also displayed.
   Most ratios from the sample of Magellanic Cloud sources show self-consistency
   but indicate significant differences when compared with
   other classes of objects.

   In analyzing molecular line intensities and line shapes,
   it is often difficult to disentangle abundance variations
   from changes in excitation conditions.
   There are, however, ways to circumvent the problem.
   By observing molecular species with similar excitation properties
   (\ie\ similar electric dipole moments and rotational constants)
   but with different chemical properties,
   changes in both chemical and physical conditions may be identified
   assuming common excitation properties.
   A good choice is a combined study of HCN, \HCOp, and
   HNC in their ground rotational $J$=1--0 transitions.
   While optical depths cannot be directly determined for most
   of these species due to a lack of hyperfine splitting, relative intensities
   provide qualitative information on relative abundances.
   \twCO, \thCO, and \CeiO\ intensity ratios are also important for an estimate
   of \twCO\ opacities and to constrain \thCO/\CeiO\ abundance ratios.

   In the following, intensity ratios between \eg\ the $J$=1--0 transitions of
   HCN and HNC will be given as $I$(HCN)/$I$(HNC), while the column density
   ratio is denoted by $N$(HCN)/$N$(HNC).

\subsubsection{\HCOp\ versus HCN}
\label{sec:MC1-hco}

   As mentioned before, \HCOp\ and HCN molecules have similar
   electric dipole moments and rotational constants.
   The observed $I$(\HCOp)/$I$(HCN) intensity ratio thus reflects directly the
   abundance ratio, \ie, $N$(\HCOp)/$N$(HCN) $>$ 1 if $I$(\HCOp)/$I$(HCN) $>$ 1
   and vice versa.
   Our observations indicate that the $I$(\HCOp)/$I$(HCN) ratios in the LMC and
   SMC are higher than in the remainder of the sample given in
   Table\,\ref{tbl:MC1-ratio} (with the notable exceptions of M82,
   \eg\ Kronberg \etal\ 1985, and the cloud associated with the SNR IC\,443).
   At low densities, \HCOp\ is produced by reactions of \Cp\ with molecules
   such as OH, \WATER, and \OO\ (Graedel \etal\ 1982).
   At high densities ($n_{\rm H_2}$ $>$ 10$^5$\,\percc), the abundance of \HCOp\
   decreases as it is lost in proton transfer reactions
   with abundant neutral molecules.
   Thus, \HCOp\ is difficult to detect in the densest cores associated with
   ultra-compact \HII\ regions (see, \eg\ Heaton \etal\ 1993).
   If much carbon is initially in the form of \Cp\
   (as is expected in the case of the Magellanic Clouds with their strong
   interstellar UV radiation fields), shocks lead to an increase in \HCOp\
   behind a shock front (Mitchell \& Deveau 1983).
   Farquhar \etal\ (1994) have discussed the effects of enhanced cosmic-ray
   fluxes on molecular abundances.
   At a given density the increase of \HCOp\ is directly
   proportional to the enhancement of the ionization.
   We thus have reason to believe that the high $I$(\HCOp)/$I$(HCN) ratios are
   caused by the intense ionization flux from supernovae,
   coupled with a large extent of the \HCOp\ emission,
   while the bulk of HCN emission arises from the dense compact cloud cores
   (see Sect.\,\ref{sec:MC1-hnc}).
   Note that such high ionization fluxes are in contrast to results derived from
   gamma-ray observations, averaged over a far larger volume than our molecular
   line observations (Chi \& Wolfendale (1993) deduced a flux of 15 $\pm$ 5\% of
   the Galactic value in the LMC and less than 11\% in the SMC).
   An analysis of the HCN hyperfine components (Sect.\,\ref{sec:MC1-hnc})
   is consistent with a confinement of HCN emission to the cloud cores.
   Combining this result with a high [C\,{\sc ii}]/CO(1--0) ratio toward the
   30\,Dor nebula, the high $I$(\HCOp)/$I$(HCN) ratios obtained from several
   Magellanic giant molecular clouds also suggest that the
   \Cp\ abundance is considerably higher than that in Galactic molecular clouds
   (\eg\ Stacey \etal\ 1991).
   This could be a result of higher UV fields and lower metallicity
   in the LMC and the SMC.

\subsubsection{HCN and HNC}
\label{sec:MC1-hnc}

   Both HCN and HNC have hyperfine structure due to the nuclear magnetic
   and quadrupole interaction from the spin of the nitrogen nucleus.
   While the HNC hyperfine components cannot be resolved, it is possible to
   resolve the HCN hyperfine components if linewidths are sufficiently narrow.
   The intensity ratios of the HCN hyperfine components, $R_{02}$
   ($F$=0--1/$F$=2--1) and $R_{12}$ ($F$=1--1/$F$=2--1) are
   known to deviate from the LTE values.
   From the observations carried out toward our Galaxy,
   $R_{12}$ has been found to be smaller in giant molecular clouds
   than the (optically thin) LTE value 0.6.
   In cold dark clouds, on the contrary, $R_{02}$ is often large relative
   to the LTE value, 0.2, in the optically thin case (see \eg\ Harju 1989).

   To explain the detected deviations from the expected LTE values of
   the HCN hyperfine ratios, Guilloteau \& Baudry (1981) have developed
   the thermal overlap model, originally presented by Gottlieb \etal\ (1975),
   which offers a reasonable explanation of the intensity ratios in warm clouds.
   According to this model, even at modest temperatures (30\,K), the overlap
   of the $J$=2--1 hyperfine transitions leads to overpopulation of the
   state $J$=1 $F$=2, and the $J$=1--0 $F$=2--1 line intensity is increased
   at the expense of the other lines.
   With increasing temperature, first the ratio $R_{12}$ and then also the
   ratio $R_{02}$ becomes smaller than the LTE values.
   In dark clouds these ratios are, on the contrary, too large as
   compared with the LTE values, and the hyperfine ratios cannot be
   explained by the thermal overlap model.

   On the other hand, Cernicharo \etal\ (1984) have considered a model
   according to which the relative intensities of the HCN hyperfine
   transitions are formed in a scattering process.
   The radiation emitted from the cloud clore is
   scattered in the surrounding envelope.
   The two optically thick lines, $F$=2--1 and $F$=1--1, are scattered more
   often than the optically thin $F$=0--1 line.
   Their emission is therefore spread over the envelope,
   while the $F$=0--1 radiation comes more directly from the core.
   Thus toward the core we should see the $F$=0--1 line enhanced relative
   to the other two hyperfine components,
   while looking slightly aside from the core the
   $F$=0--1 line will become too weak.
   Far away from the core, in the envelope, LTE intensity ratios
   are expected, if HCN is still strong enough to be observed.

   Our data allow to disentangle the HCN hyperfine structure
   toward N113 and N159HW.
   Toward these two sources $R_{12}$ is 0.61 $\pm$ 0.18 and
   0.54 $\pm$ 0.14, respectively.
   These values are close to the LTE value, 0.6.
   However, $R_{02}$ is 0.41 $\pm$ 0.16 and 0.50 $\pm$ 0.14, respectively,
   which is obviously larger than the optically thin LTE value 0.2.
   This indicates that the HCN opacity is small
   (high optical depth would reduce, not increase line intensity ratios).
   The measured ratios can be explained with the model presented by
   Cernicharo \etal\ (1984), \ie, HCN is probably emitted from
   the relative cool center of the molecular clouds.
   This interpretation is consistent with the observed $I$(\HCOp)/$I$(HCN)
   ratios (see Sect.\,\ref{sec:MC1-hco}).

   Interstellar $I$(HCN)/$I$(HNC) intensity ratios cover a range
   of at least two orders of magnitude in the Galactic disk.
   The $I$(HCN)/$I$(HNC) ratio directly reflects
   the abundance ratio because electric dipole
   moments and rotational constants of these two species are similar.
   In quiescent cool dark clouds the abundance ratio is close to, or less than,
   unity (\eg\ Churchwell \etal\ 1984; Harju 1989).
   In spiral arm gas clouds
   (Nyman \& Millar 1989), in the shocked gas associated with IC\,443
   (Ziurys \etal\ 1989), and in
   warm giant molecular clouds near sites of massive stars formation,
   (\eg\ Goldsmith \etal\ 1981, 1986; Schilke \etal\ 1992),
   the ratio increases to values of 2 -- 100.
   We find that the $I$(HCN)/$I$(HNC) intensity ratio toward
   \HII\ regions in the LMC are all larger than 2.
   Thus values $>$ 1 for the Galactic \HII\ regions
   also hold for the giant star forming regions of the LMC.

\subsubsection{$^{12}$CO, $^{13}$CO, and C$^{18}$O}
\label{sec:MC1-c18o}

   In N159 the $I$(\thCO)/$I$(\CeiO) and $I$(\CeiO)/$I$(\CseO) ratios
   observed by Johansson \etal\ (1994) indicate peculiar isotope ratios
   not found in the Galactic ISM.
   While we did not detect \CeiO\ and \CseO, our $I$(\thCO)/$I$(\CeiO)
   limits (Table\,\ref{tbl:MC1-ratio}) are even more extreme
   than the value derived by Johansson \etal\ (1994) for N159
   surpassing typical Galactic intensity ratios by a factor of 5.
   This, combined with the very low $I$(\CeiO)/$I$(\CseO) ratio of 2.0
   reported from N159 (Johansson \etal\ 1994), indicates an underabundance of
   \CeiO\ relative to \thCO\ and \CseO, not only for N159
   (Johansson \etal\ 1994) but for other \HII\ regions in the LMC as well.
   An underabundance of \CeiO\ suggests that the interstellar medium of the
   Magellanic Clouds is, relative to the Galaxy,
   dominated by low mass ($<$ 8\,\Msol) star ejecta
   (cf.\ Henkel \& Mauersberger 1993; Henkel \etal\ 1994).
   As a consequence, a `normal' \ISOC\ isotope ratio is expected,
   in spite of the low metallicity of the Magellanic Clouds
   (which would, by Galactic studies, imply \ISOC\ $\ga$ 100).
   This is consistent with Johansson \etal\ (1994) who estimate
   \ISOC\ $\approx$ 50.

\subsection{Model calculations}
\label{sec:MC1-model}

   In order to test some of the conclusions reached above, we have made some
   pseudo-time-dependent chemical kinetic calculations for molecular clouds
   with physical parameters appropriate for those in the LMC.
   We have used the model L2 of Millar \& Herbst (1990) for the
   initial abundances of C, N and O, and adopted a low dust-to-gas ratio,
   consistent with the low metallicity of the LMC.
   We have considered cloud densities in the range $3 \times 10^3$
   to 10$^4$ cm$^{-3}$ and taken a kinetic temperature of 30K.
   We have varied the cosmic-ray ionization rate between 1 and 100 times
   the Galactic average, and the interstellar UV radiation field
   between 1 and 10 times the Galactic average.
   The chemical model, which contains 187 species and 2025 reactions,
   has been extensively revised since the original LMC calculations of
   Millar \& Herbst (1990), and now includes ion-dipolar rate coefficients,
   photoreactions due to the generation of UV photons by the interaction of
   cosmic rays and \MOLH, and the latest data on neutral-neutral reactions
   (Herbst \etal\ 1994).
   Models with cosmic-ray fluxes greater than 10 times the Galactic average
   can be ruled out because they generate enough UV photons to
   destroy molecules efficiently with the result
   that calculated abundances fall below those observed.
   The chemistry is less sensitive to the size of the external UV radiation
   field, since these photons are extinguished by dust in the cloud cores.

   Table\,\ref{tbl:MC1-model} presents the abundances ratios of
   various molecules for our best-fit calculation.
   The \HCOp/HCN and CS/CN ratios are in good agreement for times
   $\approx$ 10$^4$ -- 10$^5$ years, while the CN/HCN abundance ratio is
   close to those observed $\approx$ 10$^5$ years.
   The \HCC/HCN ratio is greater than one
   at these times while the observed ratio is less than one.
   The \HCC\ abundance is very sensitive to the rate coefficient adopted for
   its major destruction reaction,
\begin{math}
     {\rm O} + \HCC\ \rightarrow {\rm CO} + {\rm CH},
\end{math}
   for which we have used $k$ = 10$^{-10}$e$^{-40/T}$\,cm$^3$s$^{-1}$.
   A rate coefficient in which the activation energy is zero would
   bring the ratios into much better agreement.
   Such a possibility is entirely reasonable.

\begin{table}
   \caption[]
           {Molecular abundance ratios calculated with our LMC model}
 \label{tbl:MC1-model}
   \begin{flushleft}
   \begin{tabular}{l r@{.}l r@{.}l r@{.}l r@{.}l r@{.}l}
   \hline
   \uspace2 \uspace-3
   Time (yrs)      & \multicolumn{2}{c}{$\frac{\HCOp}{\hbox{HCN}}$}
                   & \multicolumn{2}{c}{$\frac{\hbox{HCN}}{\hbox{HNC}}$}
                   & \multicolumn{2}{c}{$\frac{\hbox{CN}}{\hbox{HCN}}$}
                   & \multicolumn{2}{c}{$\frac{\HCC}{\HCOp}$}
                   & \multicolumn{2}{c}{$\frac{\hbox{CS}}{\hbox{CN}}$} \\
   \hline
   \uspace1
   $10^4$             &   6&8  &   1&2  &  13&1  &   7&6  &   3&2  \\
   $3.16 \times 10^4$ &   3&0  &   1&1  &   5&1  &   6&3  &   4&1  \\
   $10^5$             &   4&5  &   1&1  &   4&3  &   3&3  &   2&8  \\
   $3.16 \times 10^5$ &  42&5  &   1&1  &   3&8  &   0&3  &  15&7  \\
   $10^8$             & 121&0  &   0&7  &   2&5  &   0&01 &   8&5  \\
\uspace1 \dspace
   Observed           & \multicolumn{2}{c}{$\approx$ 2--3}
                             & $>$ 2&0  &   2&0  & $<$ 1&0
                                        & \multicolumn{2}{c}{$\approx$ 1--4} \\
   \hline
   \end{tabular}
   \end{flushleft}
\end{table}

   The HCN/HNC abundance ratio calculated is a factor 2 -- 4 less than
   the observed line intensity ratio and is very close to one.
   HCN/HNC ratios greater than 1 generally arise in warm regions (\eg\ IC\,443),
   or in spiral arm gas clouds (see Table\,\ref{tbl:MC1-ratio}).
   In both cases, the key factor is that HCN is able to form at lower
   extinctions than HNC, although at high temperatures ($T$ $>$ 100K),
   HNC may be destroyed more rapidly than HCN.
   Giant molecular clouds in the LMC and the SMC have relatively
   larger envelope/core sizes than Galactic clouds because their higher
   UV fluxes and lower metallicities allow UV photons to penetrate
   deeper into clouds (Maloney \& Black 1988).
   Under such conditions, CH$_n$ radicals form more efficiently than NH$_n$
   radicals, while \Cp, not atomic C, is the dominant form of carbon.
   Thus the reaction
\begin{math}
     {\rm N} + {\rm CH_n} \rightarrow {\rm HCN} + {\rm H_{n-1}}
\end{math}
   produces HCN faster than the reaction
\begin{math}
     {\rm C} + {\rm NH_n} \rightarrow {\rm HNC} + {\rm H_{n-1}}
\end{math}
   produces HNC.

   Our model calculations are capable only of describing the
   core of the clouds in the LMC.
   We thus predict that HCN/HNC ratios closer to one should be obtained
   if one observes at higher frequency,
   and therefore at higher spatial resolution.

\section{Conclusions}
\label{sec:MC1-Conclusions}

   From an investigation of different molecular species toward several
   \HII\ regions in both the Large and the Small Magellanic Cloud,
   we obtain the following main results:

\begin{enumerate} \renewcommand{\labelenumi}{(\arabic{enumi})}
   \item Toward N113 in the LMC we obtain the strongest
         \twCO\ and \thCO\ emission so far observed from extragalactic sources
         on a 45\arcsec\ scale.
         It is therefore recommended to carry out a detailed investigation
         of this \HII\ region.

   \item The \MOLH\ mass-to-CO luminosity conversion factors
         are $1.8\times 10^{20}$ and
         $2.4\times 10^{20}$\,mol\,cm$^{-2}$\,(K\,km\,s$^{-1}$)$^{-1}$,
         for N113 and N44BC, respectively,
         which are very close to the standard Galactic disk value but
         smaller than $X_{\rm LMC}$ obtained by the Cohen \etal\ (1988) study
         made with a much larger beam.

   \item The \thCO\ maps of observed \HII\ regions were used to estimate
         the gravitational stability.
         Ignoring magnetic fields, all cores appear to be gravitationally
         unstable.
         Disordered `irregular' magnetic fields of 30\,$\mu$G to 120\,$\mu$G
         would, however, achieve cloud stability.
         Such values are consistent with observations
         from Galactic star-forming regions.

   \item For the first time, the hyperfine components of HCN have been
         resolved for an extragalactic source.
         The normal $R_{12}$ but high $R_{02}$ values toward N113 and N159HW
         in the LMC indicate that the HCN line is emitted from a
         relatively cool dense molecular core.

   \item Some of the CN hyperfine components could be identified toward
         N113 and N159HW.
         The intensity ratios are close to the LTE values.

   \item In the Magellanic Clouds the $I$(\HCOp)/$I$(HCN) intensity ratios are
         higher than in most other galaxies (including our Galaxy).
         This ratio is even higher in the SMC than in the LMC.
         The relatively strong \HCOp\ emission is consistent with a
         high ionization flux from supernova remnants and young stars,
         while HCN is mainly arising from spatially confined dense cloud cores
         and may be optically thin.

   \item In~~~~all~~~~~studied~~~~sources~~~~of~~~~~the~~~~~LMC,
         the $I$(HCN)/$I$(HNC) ratio is $>$ 1 which is consistent with the
         presence of warm (\Tkin\ $>$ 10\,K) or shocked gas.
         Chemical model calculations suggest $I$(HCN)/$I$(HNC)
         line intensity ratios $\approx$ 1, when observed with higher
         ($<$ 30\arcsec) angular resolution.
         Only upper limits to HCN and HNC could be obtained from the SMC.

   \item The molecular hot spot N159HW has also been seen in
         radio continuum but not in \Halpha\ emission.
         This may be due to the fact that it is a relatively
         cold unevolved cloud core,
         which is consistent with the HCN hyperfine component measurements.

   \item An underabundance of \O18 \ relative to \C13 ,
         suggested by Johansson \etal\ (1994) for N159,
         is qualitatively confirmed for several star-forming regions.
         Apparently, this is a characteristic property of all
         \HII\ regions in the LMC.
\end{enumerate}

\begin{acknowledgements}
   YNC like to thank the financial support by DAAD
   (German Academic Exchange Service).
   TJM is supported by a grant from PPARC.
\end{acknowledgements}


\begin{thebibliography}{999}
 \bibitem{xyz}
   Baudry, A., Combes, F., Perault M., Dickman, R., 1980, \AaA\ \vol{85}, 244
 \bibitem{xyz}
   Bronfman, L., Cohen, R.S., Alvarez, H., May, J., Thaddeus, P., 1988,
     \ApJ\ \vol{324}, 248
 \bibitem{xyz}
   Cernicharo, J., Castets, A., Duvert, G., Guilloteau, S., 1984,
     \AaA\ \vol{139}, L13
 \bibitem{xyz}
   Chi, X., Wolfendale, A.W., 1993, J. Phys. G. \vol{19}, 795
 \bibitem{xyz}
   Churchwell, E., Nash, A.G., Walmsley, C.M., 1984, \ApJ\ \vol{287}, 681
 \bibitem{xyz}
   Cohen, R.S., Dame, T.M., Garay, G., Montani, J., Rubio, M., Thaddeus, P.,
     1988, \ApJ\ \vol{331}, L95
 \bibitem{xyz}
   Dame, T.M., Ungerechts, H., Cohen, R.S., de Geus, E.J., Grenier, I.A.,
     May, J., Murphy, D.C., Nyman, L.-\AA, Thaddeus, P., 1987,
     \ApJ\ \vol{322}, 706
 \bibitem{xyz}
   Farquhar, P.R.A., Millar, T.J., Herbst, E., 1994, \MNRAS\ \vol{269}, 641
 \bibitem{xyz}
   Fiebig, D., G\"usten, R., 1989, \AaA\ \vol{214}, 333
 \bibitem{xyz}
   Garay, G., Rubio, M., Ram\'irez, S., Johansson, L.E.B., Thaddeus, P., 1993,
     \AaA\ \vol{274}, 743
 \bibitem{xyz}
   Goldsmith, P.F., Langer, W.D., Elld\'er, J., Irvine, W., Kollberg, E.,
     1981, \ApJ\ \vol{249}, 524
 \bibitem{xyz}
   Goldsmith, P.F., Irvine, W., Hjalmarson, \AA, Elld\'er, J., 1986,
     \ApJ\ \vol{310}, 383
 \bibitem{xyz}
   Gottlieb, C.A., Lada, C.J., Gottlieb, E.W., Lilley, A.E., Litvak, M.M.,
     1975, \ApJ\ \vol{202}, 655
 \bibitem{xyz}
   Graedel, T.E., Langer, W.D., Frerking, M.A., 1982, \ApJS\ \vol{48}, 321
 \bibitem{xyz}
   Guilloteau, S., Baudry, A., 1981, \AaA\ \vol{97}, 213
 \bibitem{xyz}
   Harju, J., 1989, \AaA\ \vol{219}, 293
 \bibitem{xyz}
   Harju, J., Walmsley, C.M., Wouterloot, J.G.A., 1992, \AaAS\ \vol{98}, 51
 \bibitem{xyz}
   Heaton, B.D., Little, L.T., Yamashita, T., Davies, S.R., Cunningham, C.R.,
     Monteiro, T.S., 1993, \AaA\ \vol{278}, 238
 \bibitem{xyz}
   Heiles, C., Goodman, A.A., McKee, C.F., Zweibel, E.G.,
     1993, {\it Protostars and Planets III}, eds. E.H. Levy, J. Lunine,
     The University of Arizona Press, Tucson, p.~279
 \bibitem{xyz}
   Henkel, C., Mauersberger, R., Schilke, P., 1988, \AaA\ \vol{201}, L23
 \bibitem{xyz}
   Henkel, C., Whiteoak, J.B., Nyman, L.-\AA, Harju, J., 1990,
     \AaA\ \vol{230}, L5
 \bibitem{xyz}
   Henkel, C., Mauersberger, R., Wiklind, T., H\"uttemeister, S., Lemme, C,
     Millar, T., 1993, \AaA\ \vol{268}, L17
 \bibitem{xyz}
   Henkel, C., Whiteoak, J.B., Mauersberger, R., 1994, \AaA\ \vol{284}, 17
 \bibitem{xyz}
   Herbst, E., Lee, H-S., Howe, D.A., Millar, T.J., 1994, \MNRAS\ \vol{268}, 335
 \bibitem{xyz}
   Hunt, M.R., Whiteoak, J.B., 1994, \PASA\ \vol{11}, 68
 \bibitem{xyz}
   H\"uttemeister, S., Henkel, C., Mauersberger, R., Brouillet, N.,
     Wiklind, T., Millar, T.J., 1995, \AaA\ \vol{295}, 571
 \bibitem{xyz}
   Israel, F.P., 1992, \AaA\ \vol{265}, 487
 \bibitem{xyz}
   Israel, F.P., Johansson, L.E.B., Lequeux, J., Booth, R.S., Nyman, L.-\AA.,
     Crane, P., Rubio, M., de Graauw, Th., Kutner, M.L., Gredel, R.,
     Boulanger, F., Garay, G., Westerlund, B., 1993, \AaA\ \vol{276}, 25
 \bibitem{xyz}
   Johansson, L.E.B.,
     1991, \IAU\ \vol{146}: {\it Dynamics of Galaxies and Their Molecular
     Cloud Distributions}, eds. F. Combes, F. Casoli, p.~1
 \bibitem{xyz}
   Johansson, L.E.B., Olofsson, H., Hjalmarson, A., Gredel, R., Black, J.H.,
     1994, \AaA\ \vol{291}, 89
 \bibitem{xyz}
   Kronberg, P.P., Biermann, P., Schwab, F.R., 1985, \ApJ\ \vol{246}, 751
 \bibitem{xyz}
   Lada, C.J., 1976, \ApJS\ \vol{32}, 603
 \bibitem{xyz}
   Lovas, F.J., 1992, \JPCRef\ \vol{21}, 181
 \bibitem{xyz}
   Lequeux, J., Le Bourlot, J., Pineau des For\^ets, G., Roueff, E.,
     Boulanger, F., Rubio, M., 1994, \AaA\ \vol{292}, 371
 \bibitem{xyz}
   MacLaren, I., Richardson, K.M., Wolfendale, A.W., 1988, \ApJ\ \vol{333}, 821
 \bibitem{xyz}
   Maloney, P., Black, J.H., 1988, \ApJ\ \vol{325}, 389
 \bibitem{xyz}
   Mauersberger, R., Henkel, C., 1989, \AaA\ \vol{223}, 79
 \bibitem{xyz}
   Millar, T.J., Herbst, E., 1990, \MNRAS\ \vol{242}, 92
 \bibitem{xyz}
   Mitchell, G.F., Deveau, T.J., 1983, \ApJ\ \vol{266}, 646
 \bibitem{xyz}
   Nguyen-Q-Rieu, Nakai, N., Jackson, J.M., 1989, \AaA\ \vol{220}, 57
 \bibitem{xyz}
   Nguyen-Q-Rieu, Jackson, J.M., Henkel, C., Truong-Bach, Mauersberger, R.,
     1992, \ApJ\ \vol{399}, 521
 \bibitem{xyz}
   Nyman, L.-\AA., Millar, T.J., 1989, \AaA\ \vol{222}, 231
 \bibitem{xyz}
   Nyman, L.-\AA., Olofsson, H., Johansson, L.E.B., Booth, R.S.,
     Carlstr\"om, U., Wolstencroft, R., 1993, \AaA\ \vol{269}, 377
 \bibitem{xyz}
   Rubio, M., Lequeux, J., Boulanger, F., Booth R.S., Garay, G.,
     de Graauw, Th., Israel, F.P., Johansson, L.E.B., Kutner, M.L.,
     Nyman, L.-\AA., 1993a, \AaA\ \vol{271}, 1
 \bibitem{xyz}
   Rubio, M., Lequeux, J., Boulanger, F., 1993b, \AaA\ \vol{271}, 9
 \bibitem{xyz}
   Sage, L.J., Isbell, D.W., 1991, \AaA\ \vol{247}, 320
 \bibitem{xyz}
   Sage, L.J., Mauersberger, R., Henkel, C., 1991, \AaA\ \vol{249}, 31
 \bibitem{xyz}
   Schilke, P., Walmsley, C.M., Pineau des For\^ets, G., Roueff, E.,
     Flower, D.R., Guilloteau, S., 1992, \AaA\ \vol{256}, 595
 \bibitem{xyz}
   Stacey, G.J., Geis, N., Genzel, R., Lugten, J.B., Poglitsch, A.,
     Sternberg, A., Townes, C.H., 1991, \ApJ\ \vol{373}, 423
 \bibitem{xyz}
   Strong, A.W., Bloemen, J.B.G.M., Dame, T.M., Grenier, I.A., Hermsen, W.,
     Lebrun, F., Nyman, L.-\AA., Pollock, A.M.T., Thaddeus, P., 1988,
     \AaA\ \vol{207}, 1
 \bibitem{xyz}
   Turner, B.E., Thaddeus, P., 1977, \ApJ\ \vol{211}, 755
 \bibitem{xyz}
   Vall\'ee, J.P., 1993, \AaA\ \vol{296}, 819
 \bibitem{xyz}
   Westerlund, B.E., 1991, \AaAR\ \vol{2}, 29
 \bibitem{xyz}
   Ziurys, L.M., Snell, R.L., Dickman, R.L., 1989, \ApJ\ \vol{341}, 857
\end{thebibliography}
\end{document}